\documentclass[aps,pre,superscriptaddress,groupedaddress]{revtex4-2} 

\usepackage{amsmath}
\usepackage{graphicx}
\usepackage{dcolumn}
\usepackage{bm}
\usepackage{amsfonts} 
\usepackage{xcolor} 
\usepackage{booktabs}

\newcommand{\QC}[1]{\textcolor{black}{#1}}

\begin{document}


\title{
Low-dimensional multiscale dynamics of intermittent reversals in turbulent Rayleigh-Bénard convection
}

\author{Qiwei Chen and C. Ricardo Constante-Amores}
\affiliation{Department of Mechanical Science and Engineering, University of Illinois, Urbana Champaign, USA}

\begin{abstract}
\QC{We investigate whether a strongly turbulent flow with intermittent large-scale reorganizations admits a compact state-space description. As a representative high-dimensional chaotic system we consider
two-dimensional Rayleigh--Bénard convection at high Rayleigh number, whose dynamics are governed by multiscale interactions and rare reversals of the large-scale circulation. We introduce a multiscale latent dynamical framework in which the temporal evolution is first decomposed into slow and fast components and each is mapped to a nonlinear low-dimensional representation that is evolved by a closed dynamical system, showing that temporal scale separation alone enables an autonomous low-dimensional description of the chaotic dynamics. }
This strategy reduces the system from an original state space dimension of $\mathcal{O}(10^5)$  to a compact 20-dimensional latent space while preserving the essential multiscale dynamics. Our model reproduces the main trends of instantaneous flow structures, Reynolds stresses, energy autocorrelations, and long-time quantities such as angular momentum and wall observables, Furthermore, a waiting time analysis of flow reversals validates the statistical alignment of model prediction and DNS results. The explicit modeling of separate slow and fast branches yields significantly improved accuracy in both short-time flow structures and long-time reversal statistics, compared to single-branch alternatives. 
\QC{These results provide evidence that intermittent turbulent dynamics can evolve on a compact manifold when their intrinsic multiscale structure is respected,
offering a route toward reduced dynamical descriptions and prediction of rare events in high-dimensional chaos.}
\end{abstract}


\maketitle

\begin{quote}
\textbf{Significance statement.} 
\QC{High-dimensional chaotic systems with multiscale interactions  are commonly assumed to resist low-dimensional description, particularly when their dynamics are governed by rare, intermittent events. We show that separating slow and fast time scales prior to nonlinear dimensionality reduction yields a closed low-dimensional latent dynamical system that preserves both the long-time statistics and the switching dynamics of the original flow. This indicates that the dynamics of the intermittent turbulent flow can be captured by a compact set of latent variables when the intrinsic
multiscale structure is taken into account. By demonstrating this on a canonical high-Rayleigh-number convection system, our work establishes multiscale latent modeling as a route toward discovering reduced state-space descriptions, forecasting rare events, and connecting data-driven methods with fundamental questions on attractor dimension and predictability in high-dimensional chaos.}
\end{quote}

\section{Introduction}
\label{sec:intro}

\QC{High-dimensional chaotic systems often exhibit multiscale dynamics in which fast fluctuations coexist with slow, intermittent evolution. This pronounced separation of time scales obscures the effective dimension of the underlying attractor and challenges the construction of  low-dimensional  models \citep{menier2025interpretable, bertram2017multi}.}
Rayleigh--Bénard convection (RBC) provides a paradigmatic example of such dynamics, in which at high Rayleigh numbers the flow remains turbulent while undergoing rare, intermittent reorganizations of the large-scale circulation, making it an ideal testbed for studying the existence of finite-dimensional autonomous descriptions \citep{ahlers2009heat, lohse2010small, chillà2012new,PhysRevFluids.4.013801, STAR2021486}.



RBC is governed by two dimensionless parameters,  the Rayleigh number and the Prandtl number. The Rayleigh number, $Ra = g \beta \Delta T H^3/\nu \kappa$, quantifies the strength of buoyancy-driven flow relative to viscous and diffusive effects, where $g$ is gravitational acceleration, $\beta$ the thermal expansion coefficient, $\Delta T$ the temperature difference across the layer of height $H$, and $\nu$, $\kappa$ are the kinematic viscosity and thermal diffusivity, respectively. As $Ra$ increases, the flow undergoes a sequence of well-characterized regimes.
\QC{At Rayleigh numbers $Ra \gtrsim 10^7$, the RBC  
enters the classical turbulent regime, in which fast, small-scale fluctuations associated with plume emission and boundary-layer activity coexist with slow, large-scale dynamics such as the large-scale circulation (LSC) and its intermittent reversals~\citep{ahlers2009heat, chillà2012new, niemela2000turbulent, 10.1063/1.4744988}.}
These regimes reflect increasingly complex flow structures and multiscale interactions, presenting challenges for both theoretical analysis and numerical simulations. The Prandtl number, defined as $Pr = \nu / \kappa$, characterizes the ratio of momentum to thermal diffusivity. It plays a central role in shaping the structure and dynamics of convection. 
For moderate values ($Pr \sim 1$), such as in water or air, momentum and heat diffuse at comparable rates, producing a balance between inertial and thermal effects that yields rich multiscale dynamical behavior~\citep{ahlers2009heat}.

As $Ra$ increases,  direct numerical simulations (DNS) become computationally expensive due to the need for higher resolution to resolve all time and length scales. This has motivated the development of data-driven ROMs that aim to learn low-dimensional representations of high-dimensional flows while retaining the key physical features. Traditional linear methods, such as proper orthogonal decomposition (POD), extract the most energetic modes from flow fields and provide an orthogonal basis for truncation~\citep{annurev:/content/journals/10.1146/annurev.fl.25.010193.002543}.  When used within a Galerkin projection framework, POD-based models can suffer from instability, especially in flows with strong nonlinearities or multiscale dynamics~\citep{LoiseauBruntonNoack+2021+279+320,Schlegel_Noack_2015}. Moreover, because POD is inherently linear, it typically requires a large number of modes to capture nonlinear features accurately, limiting its efficiency compared to modern nonlinear, data-driven approaches. For example, in plane Couette flow, over 2000 degrees of freedom~\citep{Linot_Graham_2023} were required to accurately capture the system dynamics. In contrast, state-of-the-art ROMs have  achieved comparable accuracy with as few as 18 dimensions ~\citep{Linot_Graham_2023},  highlighting the efficiency and expressiveness of nonlinear latent representations.

To overcome the limitations of classical ROMs,
several manifold-based data-driven frameworks have emerged. One such method is the `Data-Driven Manifold Dynamics' (DManD) 
framework~\citep{Linot_Graham_2023}, which takes advantage of the dissipative nature of the Navier–Stokes equations (NSE), where long-term trajectories lie in a finite-dimensional attractor~\citep{zelik2022attractors}. DManD learns a coordinate transformation from the full state to a low dimensional representation using only  trajectory data, and has been proved to work well in turbulent systems such as  Couette, pipe and Kolmogorov  flow \citep{Linot_Graham_2023,Constante-Amores_Graham_2024, constanteamores2024dynamicsdatadrivenlowdimensionalmodel,disdmand,PhysRevFluids.8.044402}.
Another popular approach is the Spectral Submanifolds (SSM), which provides a mathematically rigorous method for constructing invariant manifolds in the neighborhood of known fixed points~\citep{haller2016nonlinear}, offering interpretability and preserving the geometric structure of the dynamics. However, its construction relies on specific parametrizations that can be ambiguous, and different choices may lead to varying approximation quality~\citep{stoychev2023failing}.

Recently, a growing number of studies have developed ROMs for RBC, ranging from traditional linear techniques to modern data-driven approaches. \QC{\citet{brown2007large} proposed a mathematical low-dimensional large-scale circulation model for turbulent RBC, suggesting that low-dimensional model can explain reversals. \citet{podvin2017precursor} proposed a 5-dimensional POD-based model for RBC. They reported that this model can capture reversal statistics well by introducing Gaussian noise to model coefficients. This work validated the efficiency of POD in RBC modeling and the essential contribution of small-scale noise.} \citet{10.1063/5.0168857} proposed a pressure-free, energy-conserving ROM for 2D RBC, achieving long-time stability via a structure-preserving POD-Galerkin approach. However, the method was only tested for moderate Rayleigh numbers ($10^4$–$6 \times 10^6$) and relies on conventional time-stepping schemes, which remain computationally expensive. \citet{floresmontoya2025galerkinreducedordermodel} systematically compared coupled and uncoupled POD-Galerkin ROMs. The coupled model, which employs a single basis for both velocity and temperature, better preserves the energy structure but exhibits stability issues at high Rayleigh numbers. In contrast, the uncoupled model uses separate bases for each variable, resulting in improved stability but potentially missing fine-scale interactions between velocity and temperature. 
\citet{vinograd2024} investigated nonlinear manifold learning via convolutional autoencoders, using separate autoencoders to extract latent representations from full field velocity and temperature  snapshots. Rather than combining them in a single joint state, the study treated them as distinct dynamical observables, learning parallel embeddings that each capture the latent structure of the system's inertial manifold. Although temporal evolution was not modeled, they quantified the minimal latent dimension required to accurately represent multiscale turbulent RBC  across a wide range of Rayleigh numbers ($10^6$ to $10^8$) with $Pr=1$. Their results show that the dimension increases sharply near the onset of turbulence, reflecting a rapid growth in the number of dimension as the system becomes more turbulent.

In this study, we develop a data-driven model of RBC in a  square cell exhibiting flow reversals at $Ra=10^8$ and $Pr=4.3$. \QC{Inspired by classical multiscale modeling strategies in turbulence and climate dynamics~\citep{majda2001mathematical}, we introduce a multiscale data-driven framework that captures both the fast and slow temporal components of the system dynamics.} This is achieved by applying a Gaussian filter to the POD coefficients, enabling a separation of time scales. These components are mapped into two decoupled latent spaces, each corresponding to a distinct characteristic timescale. The model accurately captures both short-time trajectory dynamics and long-time statistical behavior, all while using a small representation of only 20 degrees of freedom. 
The remainder of this article is organized as follows: Section \ref{sec:Framework} details the numerical simulation framework and outlines the methodology for identifying the low-dimensional representation and learning the associated dynamics.
Section \ref{results} presents the results related to both short-time reconstructions and long-time statistics. Finally, concluding remarks are provided in Section \ref{conclusion}.

\section{Framework}
\label{sec:Framework}

\subsection{Direct numerical simulation of RBC}

\begin{figure}
    \centering
    \begin{tabular}[t]{@{}c@{}c@{}}
        \includegraphics[width=0.4\textwidth]{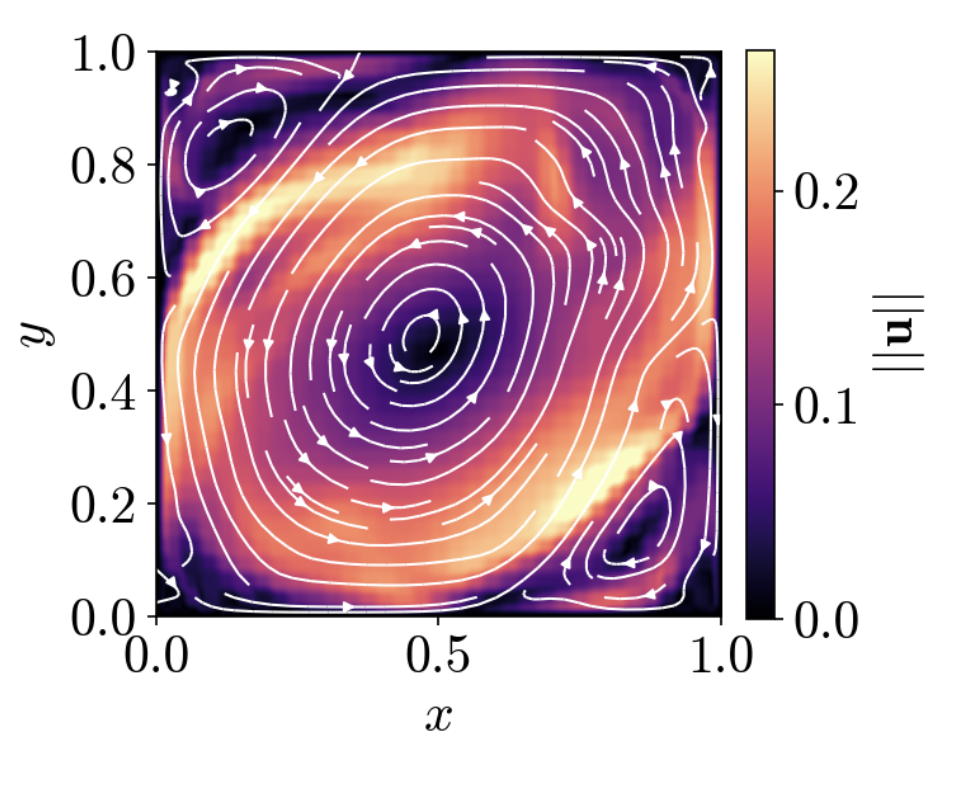} &
        \includegraphics[width=0.65\textwidth]{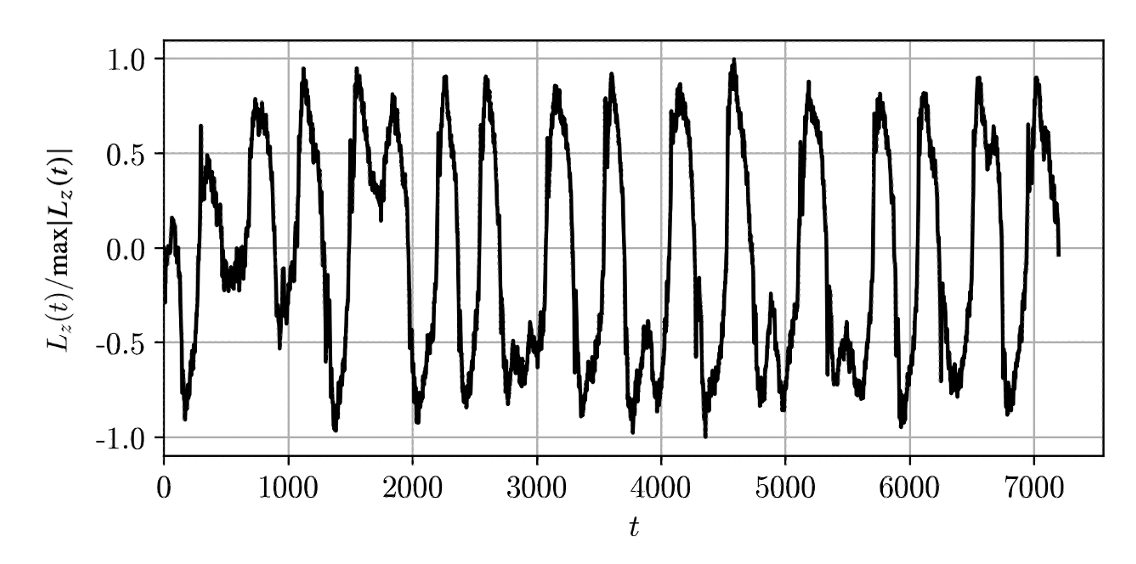} \\
        (a) & (b) \\
    \end{tabular}
    \caption{(a) Velocity field at a random instant for $Ra=10^8$ and $Pr=4.3$. Streamlines of the velocity field $\bf{u}$ are superposed over the color map.
    (b) Time evolution of  angular momentum $L_z(t)/\text{max}|L_z(t)|$.}
    \label{fig:intro}
\end{figure}

We perform  DNS of two-dimensional Rayleigh--B\'enard convection using the spectral element solver {Nek5000}~\citep{nek5000-web-page,KOOIJ201526,KOOIJ20181}. The code solves the incompressible Navier--Stokes equations coupled with the temperature transport equation under the Boussinesq approximation. The system is non-dimensionalized using the domain height $H=1$ as the characteristic length scale, the velocity scale $U = \sqrt{\Delta T H}$ based on buoyancy forces, and the advective time scale $H/U$ as the characteristic time unit. The resulting dimensionless governing equations are:
\begin{align}
\nabla \cdot \boldsymbol{u} &= 0, \\
\frac{\partial \boldsymbol{u}}{\partial t} + (\boldsymbol{u} \cdot \nabla) \boldsymbol{u} &= -\nabla p + \sqrt{\frac{\text{Pr}}{\text{Ra}}} \nabla^2 \boldsymbol{u} + T \hat{\boldsymbol{z}}, \\
\frac{\partial T}{\partial t} + (\boldsymbol{u} \cdot \nabla) T &= \frac{1}{\sqrt{\text{Ra} \cdot \text{Pr}}} \nabla^2 T,
\end{align}
where $\boldsymbol{u} = (u_x, u_y)$ is the velocity field, $T$ is the temperature, and $p$ is the pressure. The unit vector $\hat{\boldsymbol{z}}$ denotes the vertical (buoyancy) direction. The thermal boundary conditions are Dirichlet: $T = 1$ at the bottom boundary ($y = 0$) and $T = 0$ at the top boundary ($y = 1$), representing a fixed temperature difference. The velocity field satisfies no-slip boundary conditions on the horizontal walls: $\boldsymbol{u} = 0$ at $y = 0$ and $y = 1$. A representative velocity field of the system is shown in figure~\ref{fig:intro}a.

The dimensionless parameters appearing in the equations are the Rayleigh number $\text{Ra}$ and the Prandtl number $\text{Pr}$.  $Ra$ characterizes the ratio of buoyancy-driven forcing to diffusive damping, while the $Pr$ quantifies the ratio of momentum diffusivity to thermal diffusivity. They are defined as
\begin{equation}
\text{Ra} = \frac{\Delta T H^3}{\nu \kappa}, \qquad \text{Pr} = \frac{\nu}{\kappa},
\end{equation}
where $\Delta T$ is the imposed temperature difference between the bottom and top boundaries, $\nu$ is the kinematic viscosity, and $\kappa$ is the thermal diffusivity.


\subsection{Validation of numerical model and parameter selection}

In this study, we choose $\text{Ra} = 10^8$ and $\text{Pr} = 4.3$, corresponding to a strongly nonlinear convection regime with sustained LSC and intermittent reversals. The selected Prandtl number is representative of water at moderate temperature, making this test case both physically realistic and computationally challenging~\citep{PhysRevFluids.5.123502, PhysRevE.84.045303}. 

To characterize the long-time dynamics of the large-scale circulation, we compute  the dimensionless vertical angular momentum:
\begin{equation}
L_z(t) = \iint x u_y - y u_x \, dx dy, \qquad \tilde{L}_z(t) = \frac{L_z(t)}{\max |L_z(t)|},
\end{equation}
this quantity has been  successfully used to characterize flow reversals. Figure~\ref{fig:intro}b shows the normalized time series $\tilde{L}_z(t)$ over the simulation horizon. The signal change reveals quasi-periodic reversals in the direction of the mean circulation, consistent with previous studies.


The computational domain is a unit square of size $[L_x,L_y] =[0,1]^2$, discretized using a spectral element mesh consisting of $32 \times 32$ elements in the $x$ and $y$ directions, respectively. Within each element, the solution is represented using 8 Gauss–Lobatto–Legendre (GLL) points per direction, resulting in a total of $N_x = N_y = 256$ grid points. The GLL points within each element follow a Chebyshev distribution, leading to point clustering near element boundaries. This non-uniform structure is particularly beneficial for resolving  boundary layers and sharp gradients near the top and bottom walls. 
This resolution is commonly adopted in reduced-order modeling studies of Rayleigh–Bénard convection (see~\citet{CAI2019562,WANG2020104747,Yang_Schmid_2025}). 
Time integration is performed with a fixed time step size $\Delta t = 10^{-3}$, and snapshots of the velocity and temperature fields are saved every $0.1$ time units. A total of $100{,}000$ snapshots are simulated starting from a single initial condition to have a statistically steady state. To ensure that the data lie on the long-term attractor, we discard the initial transient and retain only the snapshots from time steps $20{,}000$ to $75{,}000$. \QC{The snapshots were split into 80\% for training and 20\% for  the test set.} 

To ensure sufficient spatial resolution for numerical convergence, we compute the time-averaged Nusselt number using five different definitions  adopted in RBC (see~\citet{andres_2016}). Each expression quantifies heat transport through the domain from a distinct physical perspective. Time averaging is performed over 1000 statistically independent snapshots, beginning from the 20,000th snapshot. We sample one snapshot every 50 time steps to minimize correlation~\citep{KOOIJ201526}.

\begin{equation}
\begin{aligned}
\mathrm{Nu}_{\mathrm{vol}}
&\equiv \mathrm{Ra}^{1/2}
\left\langle u_y \theta \right\rangle_{\mathrm{vol}}
- \left\langle \partial_y \theta \right\rangle_{\mathrm{vol}}, \\[4pt]
\mathrm{Nu}_{\mathrm{top}}
&\equiv - \left. \int \partial_y \theta \, \mathrm{d}x \right|_{y=1}, \\[4pt]
\mathrm{Nu}_{\mathrm{bot}}
&\equiv - \left. \int \partial_y \theta \, \mathrm{d}x \right|_{y=0}, \\[4pt]
\overline{\mathrm{Nu}}_{\theta}
&\equiv \left\langle \nabla \theta \cdot \nabla \theta \right\rangle_{\mathrm{vol}}, \\[4pt]
\overline{\mathrm{Nu}}_{\epsilon}
&\equiv \bar{\epsilon} + 1 .
\end{aligned}
\end{equation}

where $u_y$ is the vertical velocity component, $\theta$ is the temperature fluctuation field, and $\bar{\epsilon}$ denotes the volume-averaged thermal dissipation rate. The angle brackets $\left\langle \cdot \right\rangle_{\text{vol}}$ represent spatial averaging over the entire domain, while $t$ denotes instantaneous values for boundary-based definitions. Here, $\text{Nu}_{\text{vol}}$ reflects the global balance between convection and conduction. $\text{Nu}_{\text{top}}$ and $\text{Nu}_{\text{bot}}$ measure the instantaneous heat flux through the top and bottom plates. $\overline{\text{Nu}}_{\theta}$ quantifies the mean thermal gradient strength. $\overline{\text{Nu}}_{\epsilon}$ relates heat transport to the thermal dissipation rate.

\begin{table}
\centering
\label{tab:nusselt}
\begin{tabular}{lc}
\hline
Definition & $\overline{\text{Nu}}$ \\
\hline
$\text{Nu}_{\text{vol}}$ & 25.6820 \\
$\text{Nu}_{\text{top}}$ & 25.6179 \\
$\text{Nu}_{\text{bot}}$ & 25.6381 \\
$\overline{\text{Nu}}_{\theta}$ & 27.0759 \\
$\overline{\text{Nu}}_{\epsilon}$ & 27.1784 \\
\hline
\end{tabular}
\caption{\label{Nu_table}Time-averaged Nusselt numbers computed from DNS using multiple definitions based on volume- and boundary-integrated quantities}

\end{table}

Consistency across them indicates good numerical accuracy and thermodynamic balance. Table~\ref{Nu_table} reports the time-averaged values of the Nusselt number computed using them. All definitions yield consistent values near 26, indicating a well-resolved thermal balance. To further verify the consistency of the simulation, we compute the maximum relative deviation among the five definitions as
\begin{equation}
\%\text{Diff} = 100 \times \frac{\max(\text{Nu}_i) - \min(\text{Nu}_i)}{\max(\text{Nu}_i)}.
\end{equation}

In our 2D DNS  at $\text{Ra} = 10^8$, the relative deviation among three Nusselt number definitions $\text{Nu}_{\text{top}}$, $\text{Nu}_{\text{bot}}$, and $\text{Nu}_{\text{vol}}$ yields a discrepancy $\text{Diff}\approx 0.19\%$. This small deviation indicates excellent consistency among the definitions and suggests good numerical convergence. For comparison, \citet{KOOIJ20181} reported  relative difference of $0.24\%–0.73\%$ for the same  definitions in their three-dimensional simulations at $\text{Ra} = 10^8$.
At high Rayleigh numbers, the flow is highly turbulent, with  intermittent thermal plumes and fine-scale structures that can  lead to small discrepancies among  Nusselt number definitions, particularly when based on different spatial or temporal averaging strategies or dissipation-based formulations. While spatial resolution can reduce this deviation, it comes with  a substantial computational cost. For instance, simulations performed by \citet{andres_2016} reported a discrepancy of approximately $1.38\%$ across all five Nusselt definitions  at a resolution of $(512)^2$ for the same values of Ra and Pr, whereas our case yields a discrepancy of $5.2\%$. However, it is important to note that such fine resolutions result in extremely high-dimensional datasets, significantly increasing both memory requirements and computational cost of extracting coherent structures via POD. Given the small discrepancy in our results and the high computational burden of further refinement, the resolution employed in the present study represents a good compromise between accuracy and computational efficiency. 

Finally, we also note that the thermal boundary layers are resolved with five points in the vertical direction, exceeding the three points suggested by \citet{Grotzbach} for adequate resolution. This choice is consistent  with the observations of \citet{verzicco2003numerical} and \citet{amati2005turbulent}, who observed that finer resolution may be necessary at high Rayleigh numbers to capture sharp thermal gradients and ensure numerical accuracy.

\subsection{Multiscale  dimension reduction}

\subsubsection{Linear reduction: POD \label{POD_section}}

Due to the  high dimension of the RBC data (approximately $N_x \times N_y \times 3 \approx \mathcal{O}(10^5)$, where the factor of three accounts for  the two velocity components  and temperature field), it is advantageous to begin the dimension reduction  with a linear step  using POD, followed by a nonlinear reduction via autoencoders. \QC{While a fully data-driven approach based on convolutional neural networks (CNNs) could, in principle, bypass the POD step, we retain it for several reasons. First, POD provides an energetically ordered basis that captures the most energetic flow structures, allowing for a physically interpretable truncation of the state space. 
 Second, by projecting the data onto a low-dimensional POD space prior to training, we reduce the input size  to the autoencoder, which improves training efficiency and numerical stability.} Finally, our framework exploits the temporal scale separation revealed in the POD coefficients, enabling separate autoencoders to more effectively learn the fast and slow dynamics of the system.

\begin{figure}
    \centering
    \includegraphics[width=1\textwidth]{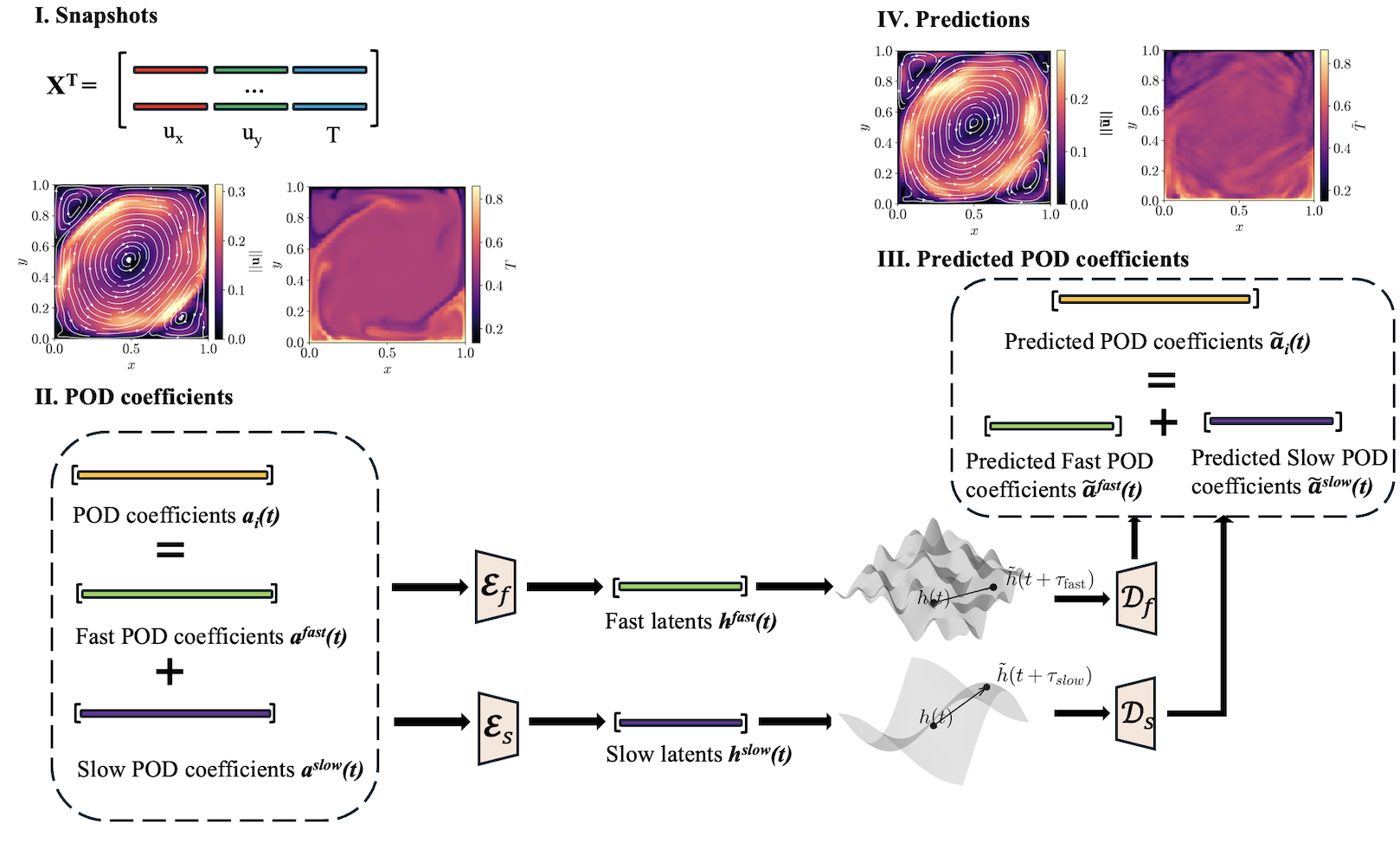}
    \caption{A cartoon of the data-driven framework applied to the full DNS of RBC. 
    }
\end{figure}



To do POD, we define the state variable as $\boldsymbol{q}(\boldsymbol{x}) = [\boldsymbol{u}(\boldsymbol{x}), T(\boldsymbol{x})]$, where $\boldsymbol{u}$ is the velocity field and $T$ is the temperature field. The aim of POD is to find a function $\boldsymbol{\Phi}$ that maximizes
\begin{equation}
    \frac{\left \langle \left |  (\boldsymbol{q}',\boldsymbol{\Phi})_E\right |^2  \right \rangle }{|| \boldsymbol{\Phi}  ||^2_E  },
\end{equation}
where $\boldsymbol{q}'(\boldsymbol{x})= \boldsymbol{q}(\boldsymbol{x}) - \bar{\boldsymbol{q}}(\boldsymbol{x})$ stands for the fluctuating component of the state variable, and $\bar{\boldsymbol{q}}$ is the mean over time. $\left \langle \cdot \right \rangle$ denotes the ensemble average, and the inner product is defined to be
\begin{equation}\label{eigenvalue1}
(\boldsymbol{q}_1,\boldsymbol{q}_2)_E = \int_\Omega \boldsymbol{q}_1 \cdot \boldsymbol{q}_2 \, d\boldsymbol{x},
\end{equation}
with the corresponding energy norm $|| \boldsymbol{q} ||_E = (\boldsymbol{q}, \boldsymbol{q})_E$. 

We apply a standard (direct) POD to the full dataset without any prior assumption of symmetries.
The POD basis functions $\boldsymbol{\Phi}^{(n)}(\boldsymbol{x})$ are obtained by solving the eigenvalue problem associated with the space-time covariance of the fluctuating state field $\boldsymbol{q}' = \boldsymbol{q} - \bar{\boldsymbol{q}}$:
\begin{equation}
\int_{\Omega} \left\langle \boldsymbol{q}'(\boldsymbol{x},t) \otimes \boldsymbol{q}'(\boldsymbol{x}',t) \right\rangle \boldsymbol{\Phi}^{(n)}(\boldsymbol{x}') \, d\boldsymbol{x}' = \lambda^{(n)} \boldsymbol{\Phi}^{(n)}(\boldsymbol{x}),
\end{equation}
where $\lambda^{(n)}$ are the POD eigenvalues corresponding to the energy captured by each mode.

To obtain these modes, we firstly organize the fluctuating state field as 
\begin{equation}
    \boldsymbol{X} = [\boldsymbol{q}'_0, \boldsymbol{q}'_1, ...,\boldsymbol{q}'_M ] \in \mathbb{R}^{N \times M}
\end{equation}
Then the empirical covariance operator is approximated as
\begin{equation}
    \mathbf{C} = \frac{1}{M - 1} X X^\top \in \mathbb{R}^{N \times N}.
\end{equation}
Direct eigendecomposition of $\mathbf{C}$ is computationally expensive for large $N$, so we apply the method of snapshots by instead solving the smaller eigenvalue problem
\begin{equation}
    \tilde{\mathbf{C}} \mathbf{v}^{(n)} = \lambda^{(n)} \mathbf{v}^{(n)}, \quad \text{where } \tilde{\mathbf{C}} = \frac{1}{M - 1} X^\top X \in \mathbb{R}^{M \times M}.
\end{equation}
The spatial POD modes are then reconstructed by projecting back to the full space:
\begin{equation}
    \boldsymbol{\Phi}^{(n)} = \frac{X \mathbf{v}^{(n)}}{\left\| X \mathbf{v}^{(n)} \right\|},
\end{equation}
where the denominator ensures $L^2$ normalization. This yields a set of orthonormal modes $\{ \boldsymbol{\Phi}^{(n)} \}$ capturing the most energetic structures in the dataset.

\subsubsection{Nonlinear reduction: Autoencoders}

After projecting the velocity and temperature fields onto the leading POD modes (i.e., in the $\mathcal{O}(10^3)$), the next step is to use autoencoders to perform a  nonlinear dimension reduction. This additional step identifies a provides a nonlinear mapping to a compact latent representation (i.e., in the $\mathcal{O}(10^1)$
). While applying autoencoders directly to full flow fields can potentially capture richer nonlinear spatial correlations, our objective here is different. The aim of the present work is not to improve spatial reconstruction beyond POD, but rather to obtain a compact nonlinear latent representation of the POD coefficients. This representation, with a much lower dimensionality than the original POD truncation, provides comparable reconstruction accuracy at substantially reduced computational cost. Training an autoencoder on the smaller set of POD coefficients is significantly more efficient than operating directly on the high-dimensional flow fields, while also retaining the energetic ordering and interpretability offered by the POD basis.

We begin by applying a Gaussian filter to the POD coefficient time series $\mathbf{a}(t)\in \mathbb{R}^{d_{POD}}$, decomposing it into slow and fast components, such as
\begin{equation}
\mathbf{a}(t) = \mathbf{a}^{\text{slow}}(t) + \mathbf{a}^{\text{fast}}(t),
\end{equation}
where $\mathbf{a}^{\text{slow}}$ captures the low-frequency, slowly evolving dynamics,  and $\mathbf{a}^{\text{fast}}$ contains the residual high-frequency fluctuations. This separation enables a more targeted analysis and modeling of the distinct temporal scales present in the system. \QC{This explicit decoupling inherently neglects direct interactions between the slow and fast components. Here, the separation is introduced as a pragmatic modeling choice rather than a claim of rigorous scale separation, with the aim of facilitating low-dimensional latent dynamics that capture both short-time evolution and long-time statistics. More rigorous closure approaches based on optimal parameterizing manifolds and stochastic corrections have shown that strong slow–fast coupling may require memory and noise terms for accurate reduced descriptions \citep{chekroun2021stochastic}.}

Next, we train two separate autoencoders: one for $\mathbf{a}^{\text{slow}}(t)$ and one for $\mathbf{a}^{\text{fast}}(t)$. Each learns a nonlinear mapping from the respective high-dimensional input to a low-dimensional latent space, $\boldsymbol{h}^{\text{slow}}(t) \in \mathbb{R}^{d_h^{\text{slow}}}, \boldsymbol{h}^{\text{fast}}(t) \in \mathbb{R}^{d_h^{\text{fast}}}$, and back. This two-branch structure enables more efficient and \QC{reasonable} modeling of the underlying dynamics by tailoring the representation to the temporal characteristics of each component.

Each autoencoder is trained to minimize the mean squared reconstruction error between the predicted and the corresponding POD coefficients.  The loss functions for the slow and fast components  are given by
\begin{align}
\mathcal{L}_{\text{slow}}(\theta_{\text{slow}}) &= \frac{1}{N} \sum_{i=1}^N \left\| \mathbf{a}^{\text{slow}}_i(t) - \mathcal{D}^{\text{slow}}(\mathcal{E}^{\text{slow}}(\mathbf{a}^{\text{slow}}_i(t))) \right\|^2_2, \\
\mathcal{L}_{\text{fast}}(\theta_{\text{fast}}) &= \frac{1}{N} \sum_{i=1}^N \left\| \mathbf{a}^{\text{fast}}_i(t) - \mathcal{D}^{\text{fast}}(\mathcal{E}^{\text{fast}}(\mathbf{a}^{\text{fast}}_i(t))) \right\|^2_2,
\end{align}
where \(\mathcal{E}^{\text{slow}}, \mathcal{D}^{\text{slow}}\) and \(\mathcal{E}^{\text{fast}}, \mathcal{D}^{\text{fast}}\) denote the encoder and decoder networks for the slow and fast components, respectively, with parameters \(\theta_{\text{slow}}\) and \(\theta_{\text{fast}}\).

Each autoencoder consists of fully connected layers, and the  neural network architectures  are summarized in Table \ref{table_arch}.
To train the autoencoders, we use  Adam optimizer to minimize the reconstruction loss 
\cite{kingma2014adam}. Each model is trained for 1000 epochs, with the learning rate decreasing from  $2 \times 10^{-3}$ to $2 \times 10^{-4}$ after epoch 500.
 Network configurations and hyperparameters were selected through an extensive trial-and-error process, in which we explored variations in depth, width, and activation functions. Our goal was to achieve accurate reconstructions while maintaining reasonable computational cost.

As we will show, this scale-separated autoencoder framework allows the model to independently  reconstruct the dominant large-scale structures and the finer-scale fluctuations, providing a flexible basis for future dynamical modeling.

\begin{table}
\centering
\begin{tabular}{cccc}
Function & Shape & Activation  & Learning Rate \\
\hline
\hline

 $ \mathcal{E}^{\text{slow}}$		& 		k/3000/500/250/$d_h^{\text{slow}}$ \quad           & GELU/GELU/GELU/lin         & $[2\times10^{-4},2\times10^{-5}]$ \\
 $ \mathcal{D}^{\text{slow}}$		& 		$d_h^{\text{slow}}$/250/500/3000/k \quad           & GELU/GELU/GELU/GELU/lin         & $[2\times10^{-4},2\times10^{-5}]$ \\
 \hline
 $ \mathcal{E}^{\text{fast}}$		& 		k/3000/1500/500/250/$d_h^{\text{fast}}$ \quad           & GELU/GELU/GELU/linear         & $[2\times10^{-4},2\times10^{-5}]$ \\
 $ \mathcal{D}^{\text{fast}}$		& 		$d_h^{\text{fast}}$/250/500/1500/3000/k \quad           & GELU/GELU/GELU/GELU/linear         & $[2\times10^{-4},2\times10^{-5}]$ \\
\end{tabular}
\caption{Neural network architectures for
slow- and fast-scales latent dynamics. `Shape' represents the dimension of each layer, `Activation'  refers to the types of activation functions used.
\label{table_arch}}
\end{table}

\subsection{NODE for slow and fast latent dynamics.}

\QC{Although alternative data-driven approaches have been applied to modeling RBC dynamics, such as reservoir computing~\citep{katsumi2025} and recurrent neural networks~\citep{Akbari09082022}, these methods are typically formulated as discrete-time models. In particular, the reservoir computing approach of \citet{katsumi2025} was trained to predict a specific integral observable, the angular momentum of the large-scale circulation, using a discrete-time mapping, rather than modeling the continuous-time evolution of the underlying flow state (i.e., velocity and temperature fields). Similarly, the RNN-based model of \citet{Akbari09082022} demonstrated good short-time predictive performance (up to approximately 25 time units), but was restricted to relatively low Rayleigh numbers (up to $10^7$) and did not report long-time dynamical behavior. In contrast, the present work aims to construct a reduced representation that admits an explicit continuous-time dynamical closure in a latent state space. To this end, we adopt NODEs to model the temporal evolution of the latent variables, enabling a compact and autonomous description of the system dynamics that is naturally suited for long-time integration and statistical analysis.}


Once the slow and fast components have been projected onto  their respective low-dimensional representations, we can define an evolution equation in the latent space, 
\begin{equation} \label{eq:Node1}
\frac{d\boldsymbol{h}}{dt} = \boldsymbol{g}(\boldsymbol{h}),
\end{equation}
where the vector field $\boldsymbol{g}(\boldsymbol{h})$ is represented by a neural network that approximates the nonlinear dynamics in latent coordinates.


To respect the intrinsic scale separation between slow and fast dynamics, we train two distinct NODE models, each  capturing the temporal evolution of one component. Then, the slow and fast latent trajectories are governed by independent NODE of the form

\begin{align}
\frac{d\boldsymbol{h}^{\text{slow}}}{dt} &= \boldsymbol{g}_{\text{slow}}(\boldsymbol{h}^{\text{slow}}), \label{eq:node_slow} \\
\frac{d\boldsymbol{h}^{\text{fast}}}{dt} &= \boldsymbol{g}_{\text{fast}}(\boldsymbol{h}^{\text{fast}}), \label{eq:node_fast}
\end{align}
where \(\boldsymbol{g}_{\text{slow}}\) and \(\boldsymbol{g}_{\text{fast}}\) are distinct neural networks in the form of equation \ref{eq:Node1}. \QC{The use of two independent NODEs reflects a modeling assumption motivated by temporal scale separation in the latent dynamics, and does not imply a strict physical decoupling or the absence of interactions between slow and fast processes in the underlying flow.}

Although the slow and fast dynamics are modeled by separate ODE systems, they are trained jointly by minimizing a unified loss function. Each system is integrated forward from time $t_i$ to $t_i + \tau$ to generate predictions $\tilde{\boldsymbol{h}}^{\text{slow}}(t_i + \tau)$ and $\tilde{\boldsymbol{h}}^{\text{fast}}(t_i + \tau)$ respectively. These latent representations are mapped back to reconstruct the predicted POD coefficients $\tilde{\boldsymbol{a}}(t_i + \tau)$.

We define the loss function as 

\[
J = \frac{1}{N} \sum_{i=1}^N \left\| \tilde{\boldsymbol{h}}^{\text{slow}}(t_i + \tau) - \boldsymbol{h}^{\text{slow}}(t_i + \tau) \right\|_2^2 +  \frac{1}{N} \sum_{i=1}^N \left\| \tilde{\boldsymbol{h}}^{\text{fast}}(t_i + \tau) - \boldsymbol{h}^{\text{fast}}(t_i + \tau) \right\|_2^2,
\]
where the first and second  terms of the RHS are the MSE for the slow and fast components of the latent state, respectively. Here,
$\tilde{\boldsymbol{h}}^{\text{slow}}(t_i + \tau)$ and $\tilde{\boldsymbol{h}}^{\text{fast}}(t_i + \tau)$ are defined as 

\[
\tilde{\boldsymbol{h}}^{\text{slow}}(t_i + \tau) = \boldsymbol{h}^{\text{slow}}(t_i) + \int_{t_i}^{t_i + \tau} \left( \boldsymbol{g}_{\text{slow}}(\boldsymbol{h}^{\text{slow}}(t); \theta_{g_{\text{slow}}}) - \boldsymbol{A}_1 \boldsymbol{h}^{\text{slow}}(t) \right) \, dt,
\]


\[
\tilde{\boldsymbol{h}}^{\text{fast}}(t_i + \tau) = \boldsymbol{h}^{\text{fast}}(t_i) + \int_{t_i}^{t_i + \tau} \left( \boldsymbol{g}_{\text{fast}}(\boldsymbol{h}^{\text{fast}}(t); \theta_{g_{\text{fast}}}) - \boldsymbol{A}_2 \boldsymbol{h}^{\text{fast}}(t) \right) \, dt,
\]
here, $ \boldsymbol{A}_1 $ and $ \boldsymbol{A}_2 $ are  linear damping terms that are trained along with the rest of the model to prevent  divergence in the latent trajectories and to stabilize the long time  dynamics (as shown in \citet{Linot_Graham_2023}). A summary of the architecture is given in the Appendix. From a physical perspective, the damping terms mimic viscous dissipation, suppressing nonphysical growth in the latent states. Without such dissipation, the learned dynamics could drift away from the attractor or exhibit unbounded energy growth. These terms serve  a role analogous to regularization in optimization.

This multi-part loss ensures that both latent representations are accurately modeled, as we will show in the next section. Gradients from all loss components are backpropagated to jointly optimize both $\boldsymbol{g}_{\text{slow}}$ and $\boldsymbol{g}_{\text{fast}}$ \citep{chen2018neural}. Training is performed using the Adam optimizer with a scheduled learning rate decay to improve convergence and stability.

\section{Results \label{results} }

\subsection{Dimension reduction}

We begin by performing POD on the flow field, as described in section \ref{POD_section}, using   the snapshots to obtain an energy-ranked orthogonal basis for the system.
The POD modes are ordered  in descending order from largest to smallest eigenvalues ($\lambda_i$).  The flow field  is then  projected onto the leading $d_{POD}=1509$ modes, which  retain  99.95\% of the total energy of the system. We have observed that including additional modes  does not significantly improve the reconstructed field or  predictive accuracy, but it does increase computational cost in the subsequent nonlinear modeling stages.

\begin{figure}
\centering
\begin{tabular}{cc}
\includegraphics[width=0.5\textwidth]{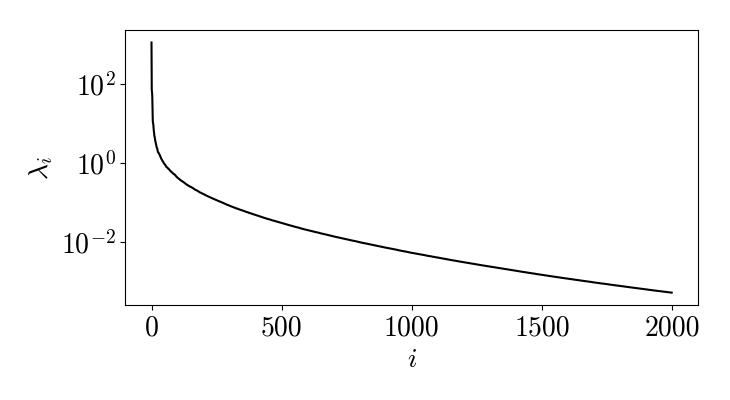}     &
\includegraphics[width=0.5\textwidth]{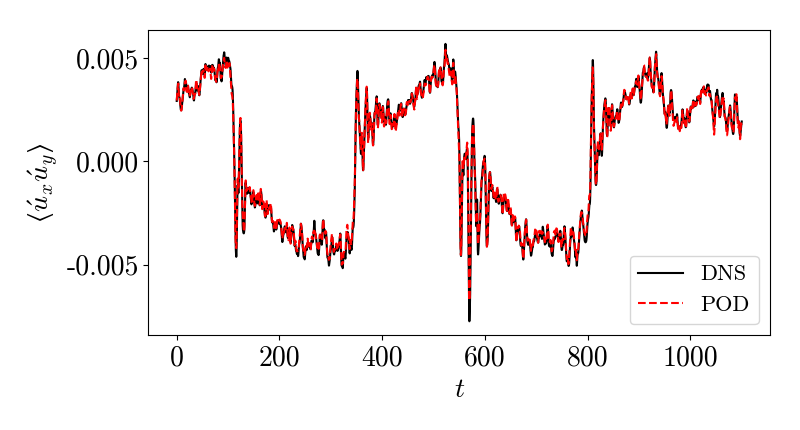} \\
(a) & (b) \\
\includegraphics[width=0.5\textwidth]{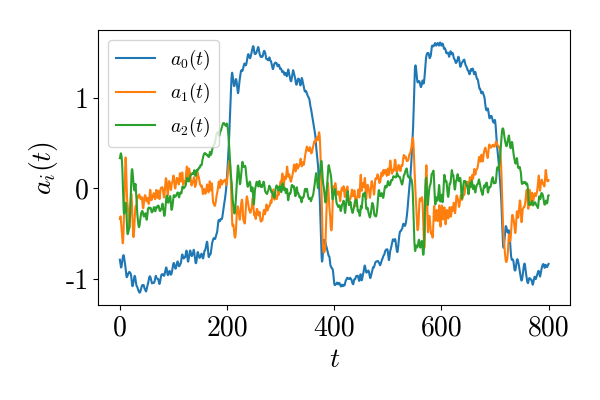}     &
\includegraphics[width=0.5\textwidth]{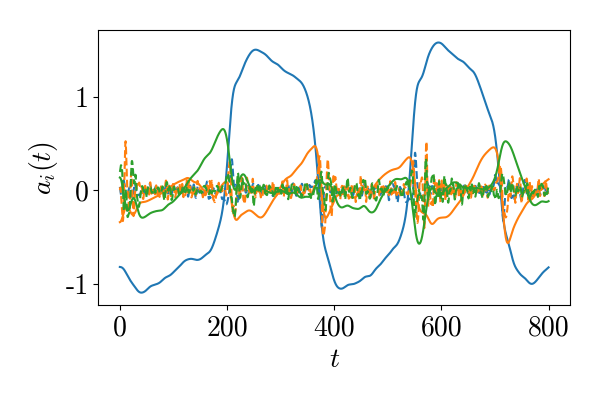} \\
(c) & (d) \\
\end{tabular}
\caption{Linear dimension reduction: (a) Spectrum of POD eigenvalues $\lambda_i$. (b) Comparison of the Reynolds stress component $\langle u_x'u_y' \rangle$ from DNS and POD reconstruction. (c)  Temporal evolution of the modal amplitudes, $a_i(t)$, for the three leading POD modes. (d) Decomposition of the first three POD coefficients into slow  and fast  components using Gaussian filtering. Solid line represents slow components while dash line represents fast components.}
    \label{fig_pod}
\end{figure}

Figure~\ref{fig_pod}a shows the eigenvalue spectrum of the POD modes, where $\lambda_i$ are normalized by the leading eigenvalue $\lambda_1$. The spectrum exhibits a rapid initial decay: for instance, the first 67 modes capture 95\% of the total energy, while 359 modes  account for 99\%. Beyond these leading modes, the eigenvalues exhibit a slow decay, suggesting a large number of energetically insignificant modes that contribute primarily to fine scale fluctuations.  Figure~\ref{fig_pod}b compares the Reynolds stress component $\langle u_x'u_y' \rangle$ computed from the full DNS data and the 1509-mode POD reconstruction. The good match between them indicates that 1509 POD modes are sufficient to accurately  preserve key second-order statistics of turbulence.

Figure~\ref{fig_pod}c shows the temporal evolution of the first three POD modal coefficients $a_i(t)$ over $800$ time units. Each mode exhibits  slow and fast temporal dynamics, highlighting  the underlying multiscale structure of the system.
To isolate the multiscale temporal features embedded in the POD coefficients, we apply a Gaussian filter along the time axis to each modal coefficient. This filtering  enables the separation of the original POD coefficient $a_i(t)$  into low-frequency (slow) and  high-frequency  (fast) components, such as
\begin{equation}
\mathbf{a}(t) = \mathbf{a}^{\text{slow}}(t) + \mathbf{a}^{\text{fast}}(t),
\end{equation}
where $\mathbf{a}^{\text{slow}}(t)$ is obtained via convolution with a Gaussian kernel:
\begin{equation}
    \mathbf{a}^{\text{slow}}(t) = \int_{-\infty}^{\infty}  \mathbf{a}(s) \exp\left(-\frac{(t-s)^2}{2\sigma^2}\right) \, ds,
\end{equation}
and the fast component is defined as the residual, $\mathbf{a}^{\text{fast}} = \mathbf{a}(t) - \mathbf{a}^{\text{slow}}(t)$. 
In our implementation, we used a standard deviation $\sigma$ for the Gaussian filter. This decomposition allows us to analyze the contributions of slow, coherent motions and fast, fluctuation-    driven dynamics separately, providing a clearer view of the hierarchical temporal structure embedded in the flow. We note, however, that this filtering is essentially an ad hoc procedure \citep{Brett2011StabilityOF}. A more rigorous decomposition could be pursued by, for example, spectral analysis of POD coefficients, wavelet-based methods, or empirical mode decomposition. A systematic comparison of such alternatives is beyond the scope of the present study, but represents an interesting direction for future work.

Figure~\ref{fig_pod}d presents the POD coefficients for the first three modes, each decomposed into  slow and fast temporal components. The slow scales components  display pronounced low‑frequency oscillations, corresponding to   large-scale flow reversals; while the fast scales components remain relatively small in amplitude, except during instances when the slow-scale components cross zero or undergo sharp directional changes.

This decomposition reveals a hierarchical coupling between temporal and spatial structures in the flow. The slow temporal components, associated with large-scale spatial structures, govern the long-term evolution of the system, including events such as flow reversals. In contrast, the fast temporal components, linked to small scale spatial fluctuations, are lower in amplitude but become  significant during transitions. This interaction highlights that, although fast fluctuations may appear as noise in some contexts, they play a critical role in modulating and sustaining the large-scale dynamics in RBC.

After projecting the full state of dimension $\mathcal{O}(10^5)$ onto a reduced basis of dimension $\mathcal{O}(10^3)$ using POD, we use autoencoders to identify a nonlinear latent representation with even fewer degrees of freedom. We first centre and scale the dataset by subtracting
the mean and then normalizing it with the maximum standard deviation to enhance training stability while preserving the relative energy significance of the POD coefficients. 

\begin{figure}
    \centering
\begin{tabular}{cc}
\includegraphics[width=0.5\textwidth]{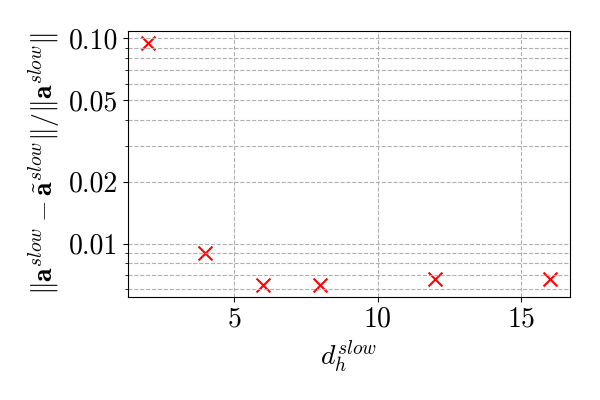}     & \includegraphics[width=0.5\textwidth]{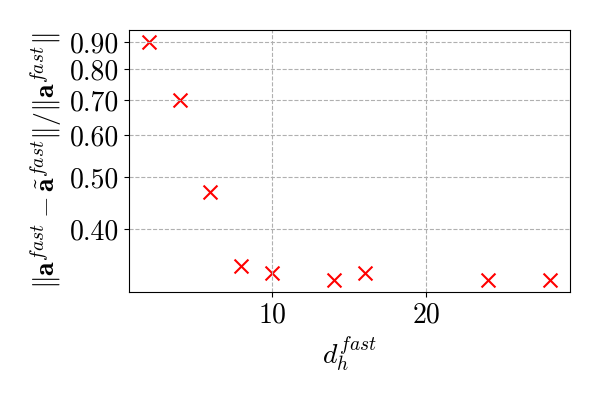} \\
 (a)    &  (b)
\end{tabular}
\begin{tabular}{c}
\includegraphics[width=0.5\textwidth]{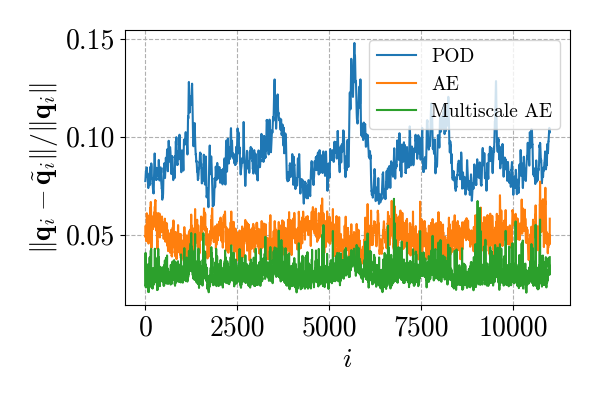} \\
 (c)  \\
\end{tabular}
\caption{Non linear dimension reduction: (a) Relative reconstruction error as a function of latent dimension $d_h^{\text{slow}}$ for slow-scale autoencoder. (b) Relative reconstruction error as a function of latent dimension $d_h^{\text{fast}}$ for fast-scale autoencoder. (c) Temporal evolution of the relative reconstruction error for POD, a single-stage autoencoder with $d_h=20$, and the proposed multiscale autoencoder with total latent dimension $d_h=20$.}
    \label{fig:err_vs_dh_slow}
\end{figure}

To determine the minimal dimension required to accurately represent the system dynamics, we train multiple autoencoders with varying latent space dimensions, and evaluate their performance by computing the relative reconstruction error on the test data set, \QC{ defined as $ E_{\text{rel}} =\| \mathbf{a} - \hat{\mathbf{a}} \|_2/\| \mathbf{a} \|_2$, where $\mathbf{a}$ denotes the original POD coefficients and $\hat{\mathbf{a}}$ its reconstruction.
}
Given the distinct temporal scales inherent in the flow, we  train two separate autoencoders: one targeting  the slow (low-frequency) components, and another for the fast (high-frequency) components of the POD coefficients.
Figures \ref{fig:err_vs_dh_slow}a,b show the relative reconstruction errors as functions of latent dimension    $d_h^{\text{slow}}$ and $d_h^{\text{fast}}$, respectively.For the slow dynamics, the relative reconstruction error decreases rapidly with increasing $d_h^{\text{slow}}$, and plateaus beyond $d_h^{\text{slow}} > 6$ at approximately $0.007$, with no appreciable improvement for higher dimensions. In contrast, the fast dynamics exhibit a steeper initial decline in relative reconstruction error as $d_h^{\text{fast}}$ increases, but the error saturates  beyond $d_h^{\text{fast}} > 10$ at a relatively higher value of about  0.32. The comparatively larger error for the fast dynamics can be attributed to their larger spectral content and weaker temporal correlation, which limit the effectiveness of low-dimensional representation. Based on these trends, we selected $d_h^{\text{slow}} = 6$ and $d_h^{\text{fast}} = 14$, yielding a combined latent dimension of $d_h = 20$ for the full system.

\QC{To assess the relative performance of different low-dimensional representations (i.e., from POD, standard AE and multiscale AE) for the same  dimension, figure ~\ref{fig:err_vs_dh_slow}c shows the  temporal evolution of the relative reconstruction error with  $d_h = 20$. Here, the reconstruction error is defined as
\[
E_{\mathrm{rel}} = \| \mathbf{q} - \hat{\mathbf{q}} \|_2 / \| \mathbf{q} \|_2,
\]
where $\mathbf{q}$ denotes the original snapshot and $\hat{\mathbf{q}}$ the reconstructed snapshot. All models shown in Fig.~\ref{fig:err_vs_dh_slow}c are compared using the same effective number of degrees of freedom. Under this fixed dimension, the multiscale autoencoder yields consistently lower reconstruction error over time compared to both a linear POD truncation and a standard single-stage autoencoder. This reduction reflects a more efficient nonlinear representation of the dynamics within the prescribed latent dimension, rather than the recovery of spatial structures beyond the POD subspace. In particular, the average relative reconstruction error decreases from approximately $0.1$ for POD to $0.03$ for the multiscale model, while the single-stage autoencoder achieves an intermediate value of approximately $0.05$. The comparison further highlights the benefit of explicitly separating slow and fast temporal components, as the multiscale architecture provides improved representation efficiency relative to a single-stage latent model at the same total dimension. We emphasize that this improvement reflects a more efficient use of a fixed low-dimensional latent space, rather than enhanced physical fidelity or the reconstruction of flow structures beyond the POD subspace. In particular, the comparison is performed against a POD truncation with the same number of modes, and no claim is made regarding performance relative to high-energy POD reconstructions (e.g., retaining 99\% of the total energy).}



\begin{figure}
    \centering
\begin{tabular}{c}
    \includegraphics[width=\textwidth]{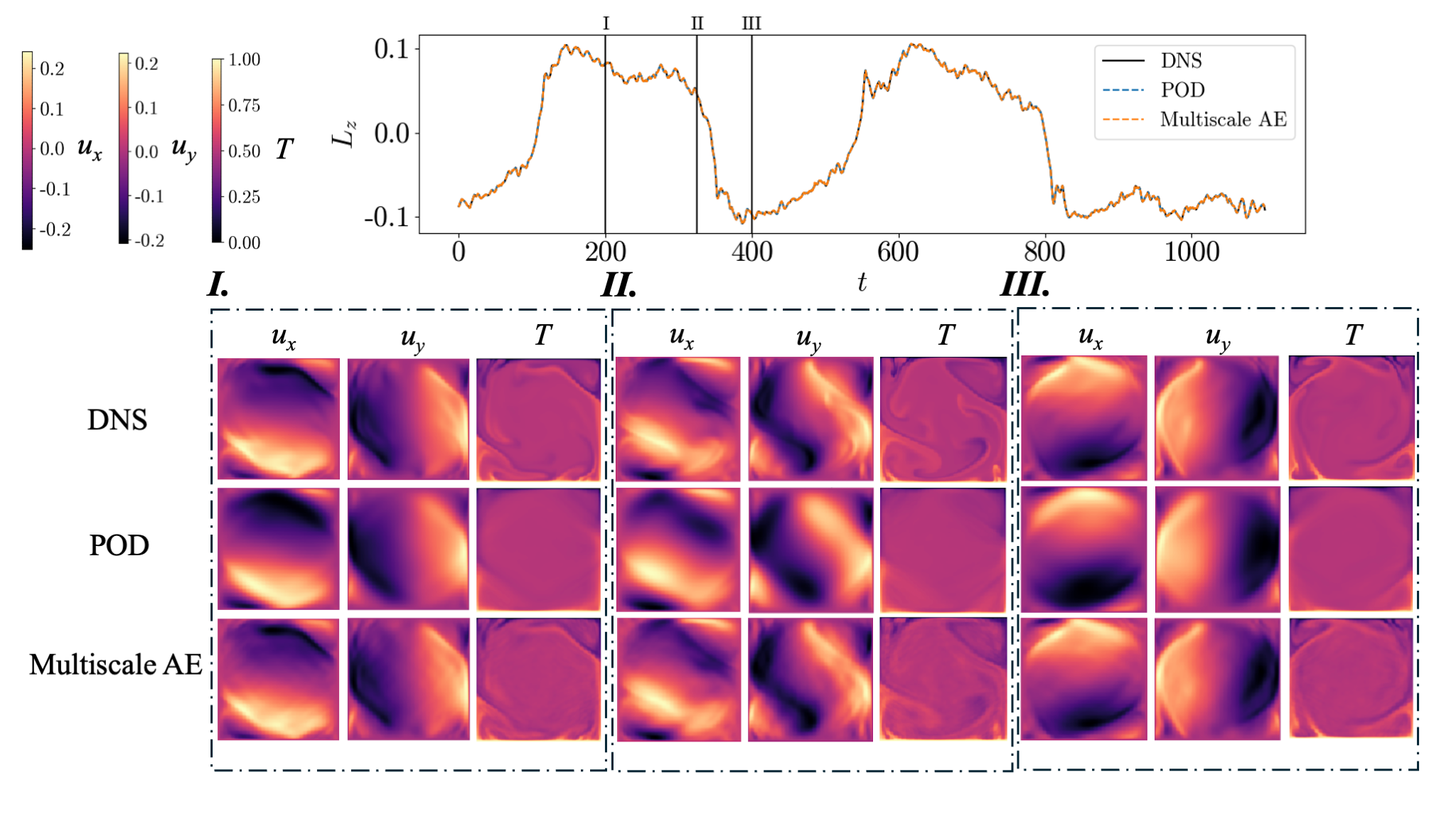}\\
    (a)
\end{tabular}
    \centering
\begin{tabular}{ccc}
\includegraphics[width=0.33\textwidth]{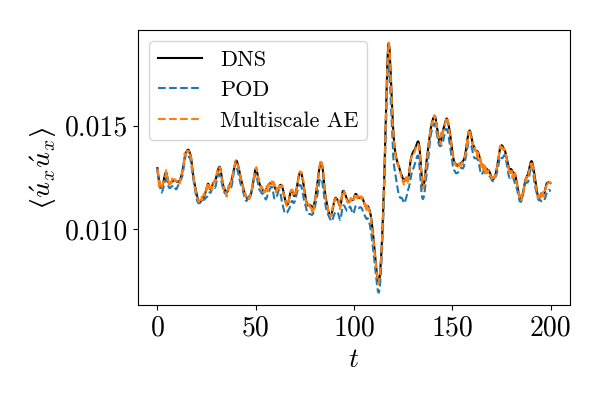}&
\includegraphics[width=0.33\textwidth]{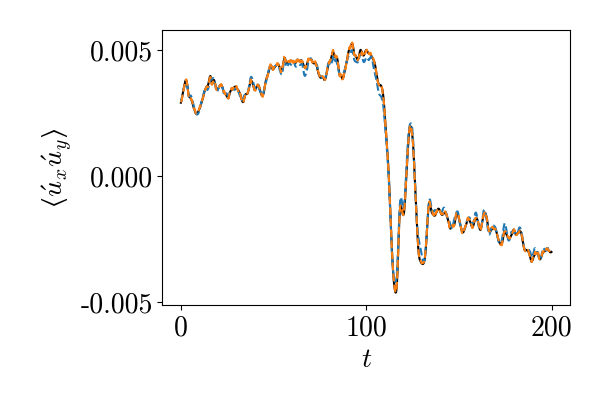}& 
\includegraphics[width=0.33\textwidth]{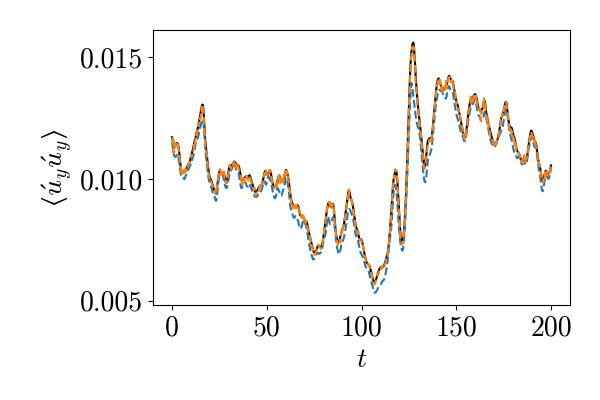}
\\
 (b)    & (c)  & (d) \\
\end{tabular}
\caption{(a) Comparison of angular momentum $L_z$ and reconstructed velocity and temperature fields at select time shown in the top panel. (b-d) Reynolds stresses of $\langle u_x'u_x' \rangle$, $\langle u_x'u_y' \rangle$  and $\langle u_y'u_y' \rangle$ among DNS, POD, and multiscale autoencoder, corresponding to panels b to d, respectively.     }
    \label{fig:lz_comparison}
\end{figure}

Figure~\ref{fig:lz_comparison} shows the temporal evolution of the angular momentum $L_z$ using DNS, POD, and the multiscale autoencoder across several reversal cycles. As a global quantity, it does not reveal a significant  improvement of the multiscale autoencoder over POD, motivating a closer examination of instantaneous flow fields (e.g., velocities and temperature fluctuations) at selected times.   We focus on  snapshots at   $t=200$, $t=325$ and $t=400$,  which correspond  to moments when the angular momentum changes sign, indicating  flow reversals and transitional dynamics in the system. 
Visual inspection shows  that the POD reconstruction  captures only the dominant  large-scale structures of the flow but fails to reproduce the small scale features, such as thermal plumes. In contrast, the multiscale autoencoder successfully reconstructs both large   and fine scale structures, owing to its ability to capture and reconstruct both slow and fast temporal dynamics through a multiscale representation. This results in a more physically faithful and dynamically consistent representation of the flow, crucial for predictive modeling.
To further assess the model's ability to capture turbulent fluctuations during flow reversals, we compute the spatially averaged Reynolds stresses components $\langle u_x'u_x' \rangle$, $\langle u_x'u_y' \rangle$, and $\langle u_y'u_y' \rangle$ as functions of time, where $\langle \cdot \rangle$ denotes averaging over the spatial domain at each time instant
(see figures~\ref{fig:lz_comparison}b-d). 
The multiscale AE yields excellent agreement with the DNS results across all components, significantly outperforming the POD based reconstruction \QC{with the same dimension}, particularly in regions with intense fluctuation activity.

\begin{figure}
    \centering
\begin{tabular}{cc}
\includegraphics[width=0.45\textwidth]{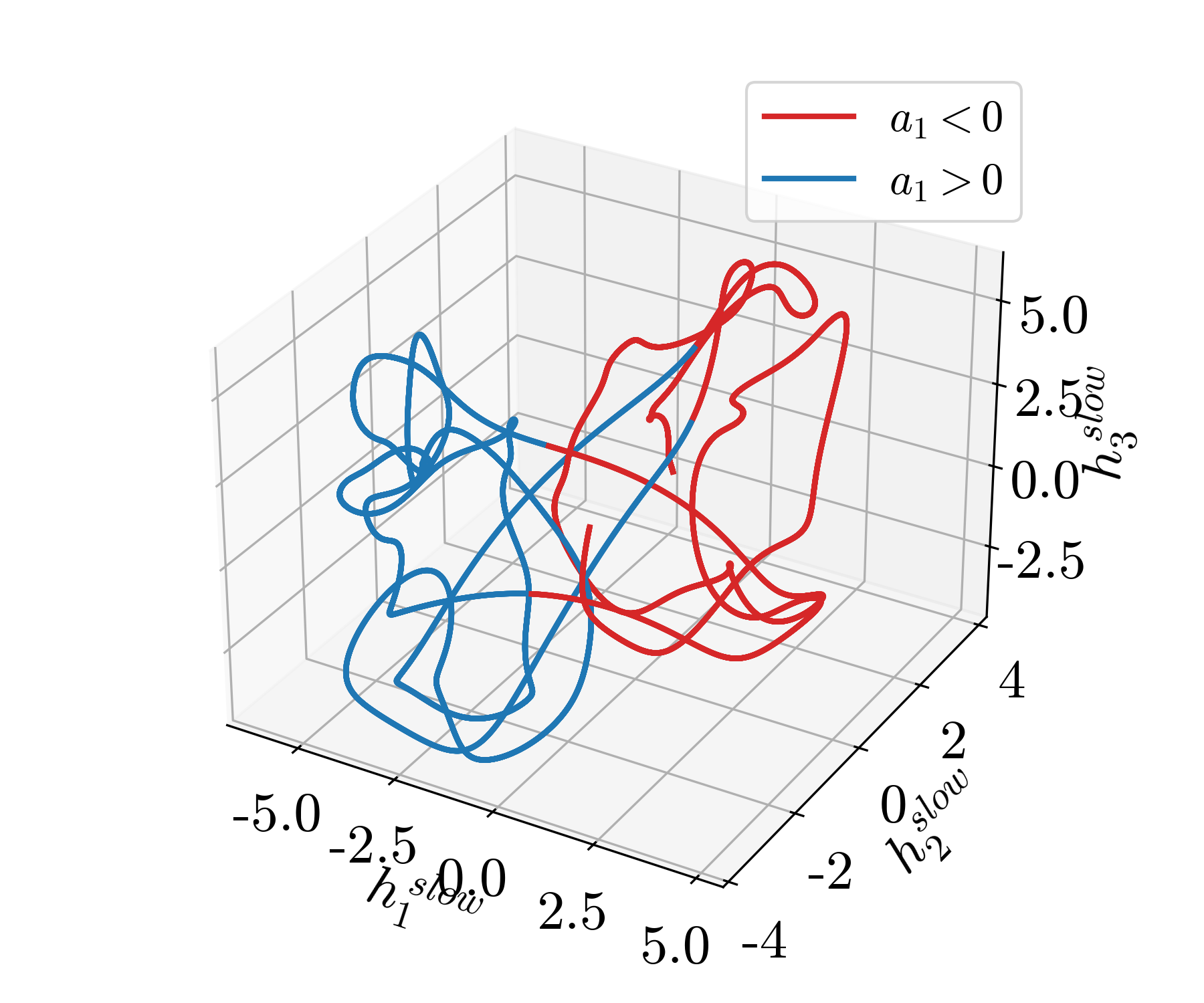}&
\includegraphics[width=0.45\textwidth]{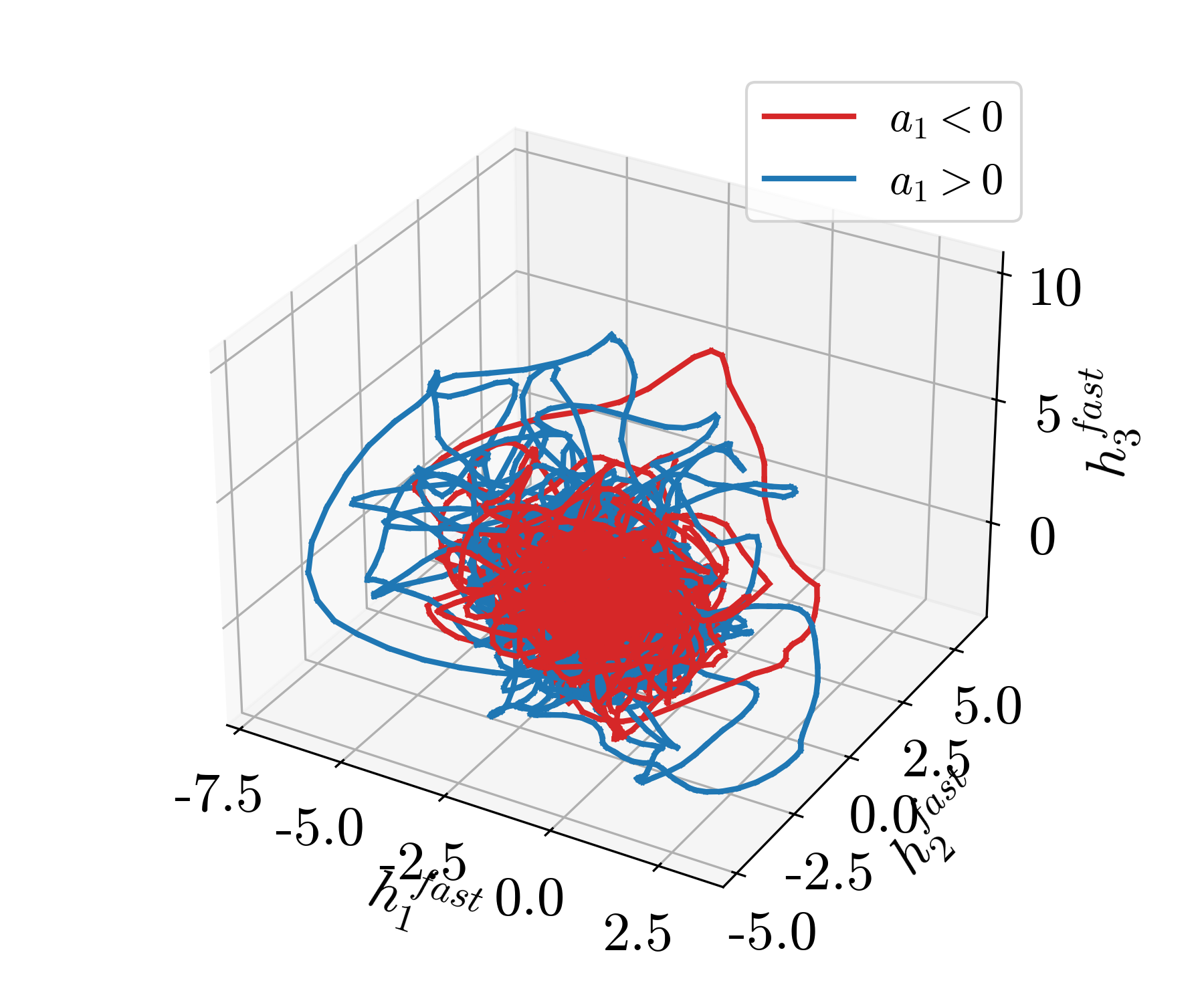}
\\
 (a)    & (b)\\
\end{tabular}
\caption{DNS trajectory of the slow (a) and fast (b) scale latent variables  for trajectory shown in figure 5. The red segment indicates the first POD coefficients $a_1 < 0$, while the blue segment indicates $a_1 > 0$.}
    \label{fig:traj}
\end{figure}

Figure \ref{fig:traj} shows the DNS trajectory projected onto  the slow and fast scale latent variables. This result reveals the physical interpretation of scale separation achieved by our multiscale decomposition. The slow scale trajectory appears well organized, and clustered according to the sign of the firs POD coefficient, $a_1$, which reflects   LSC reversal (e.g., where $a_1>0$ and $a_1<0$ correspond to clockwise and counterclockwise rotation, respectively). Two distinct regions in latent space are observed, corresponding to opposite LSC directions, consistent with previous observations of bistability in the system. In contrast, the fast-scale trajectory is more chaotic and unstructured, consistent with the turbulent nature of the underlying dynamics. \QC{This separation offers a possible interpretation of the latent variables. The slow-scale component encodes the bistable LSC orientation, while the fast-scale component captures turbulent variability.} 

Finally, we note that this three-stage reduction, from the full state space of dimension $\mathcal{O}(10^5)$, to a POD-reduced representation of dimension $\mathcal{O}(10^3)$, and finally to a nonlinear latent space of dimension $\mathcal{O}(10^1)$, demonstrates the effectiveness of the proposed framework in capturing the essential dynamics with a reduction of the dimension by four orders of magnitude.


\subsection{Time evolution: NODE models}

After identifying  the  minimal dimension required to capture the slow and fast dynamics, we proceed to learn a model in this low dimensional representation using NODE \citep{chen2018neural,LINOT2023111838}.  Prior to  training, the latent trajectories  are centered by subtracting their temporal mean. This step is essential to prevent the learned dynamics from drifting away from the low dimensional representation and to ensure long time stability and consistency with the dissipative nature of the original high dimensional system.

Throughout all comparisons between the DNS and the data-driven models, identical initial conditions are employed. For each model, the  initial condition is first projected onto the low dimensional representation, and then evolved forward in time using the appropriate NODE. In the case of the standard NODE, this evolution generates a trajectory for $(\boldsymbol{h}(t))$, whereas for the multiscale NODE, separate trajectories for $\boldsymbol{h}^{\text{slow}}(t)$ and $\boldsymbol{h}^{\text{fast}}(t)$ are obtained. In both cases, the latent variables are subsequently mapped back to the full space to reconstruct the physical fields for direct comparison against the DNS.


\begin{figure}
    \centering
    \begin{tabular}{c}
        \includegraphics[width=0.85\textwidth]{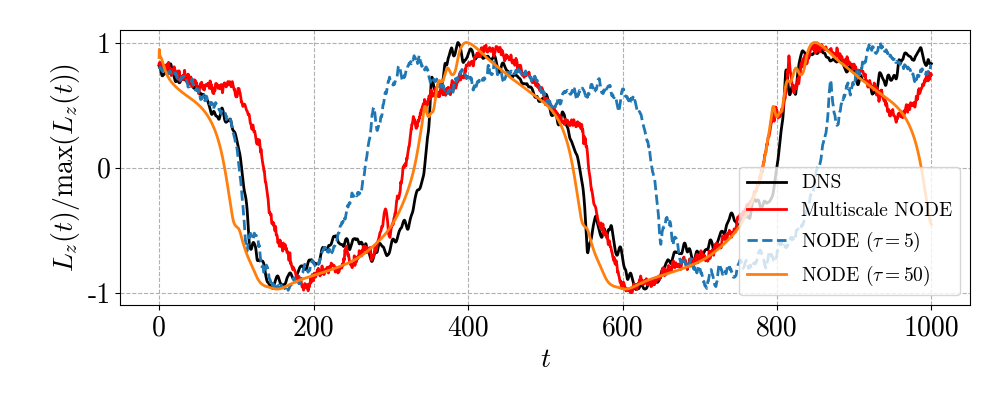} \\
        (a)
    \end{tabular}
    \begin{tabular}{cc}
        \includegraphics[width=0.42\textwidth]{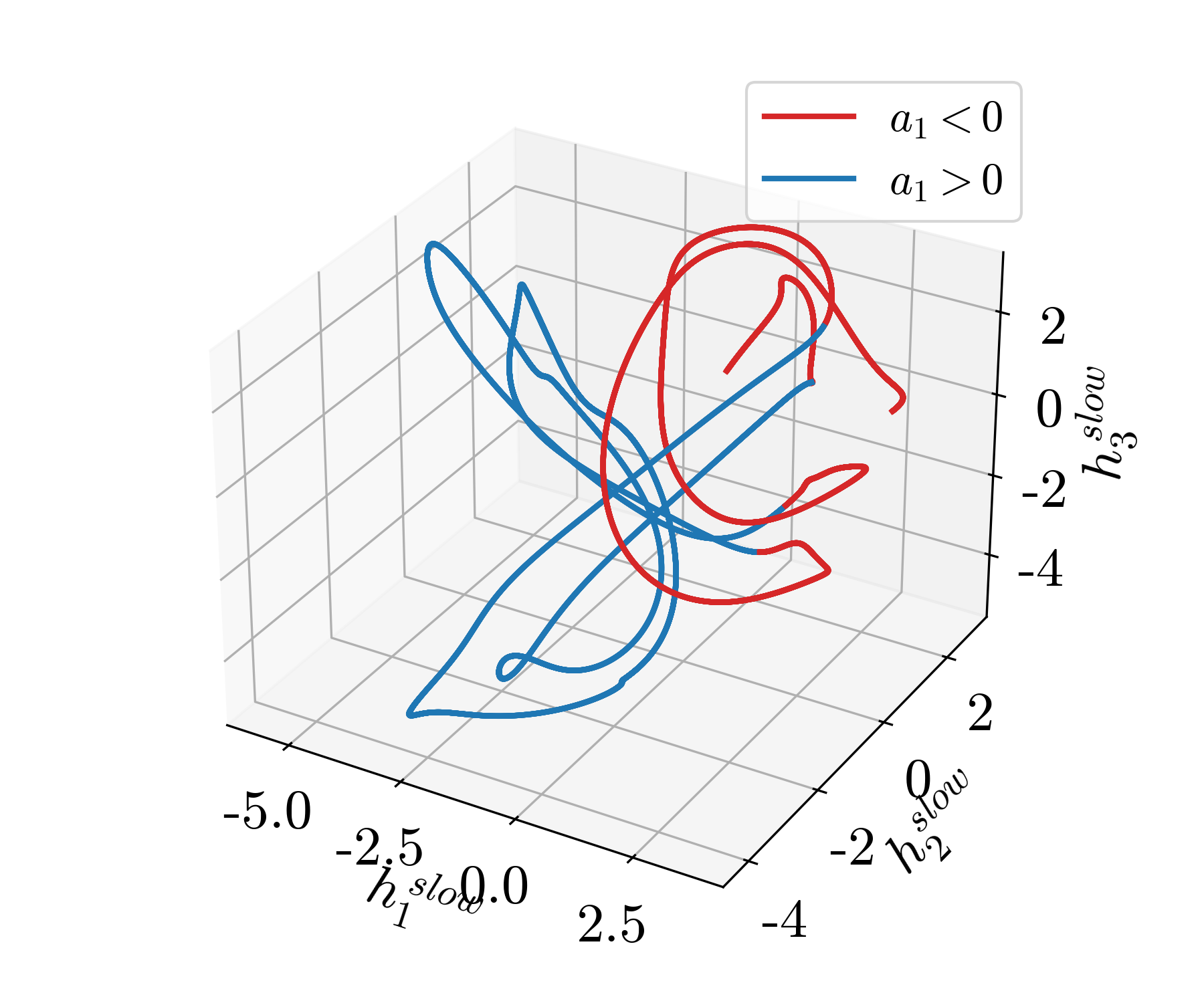} &
        \includegraphics[width=0.42\textwidth]{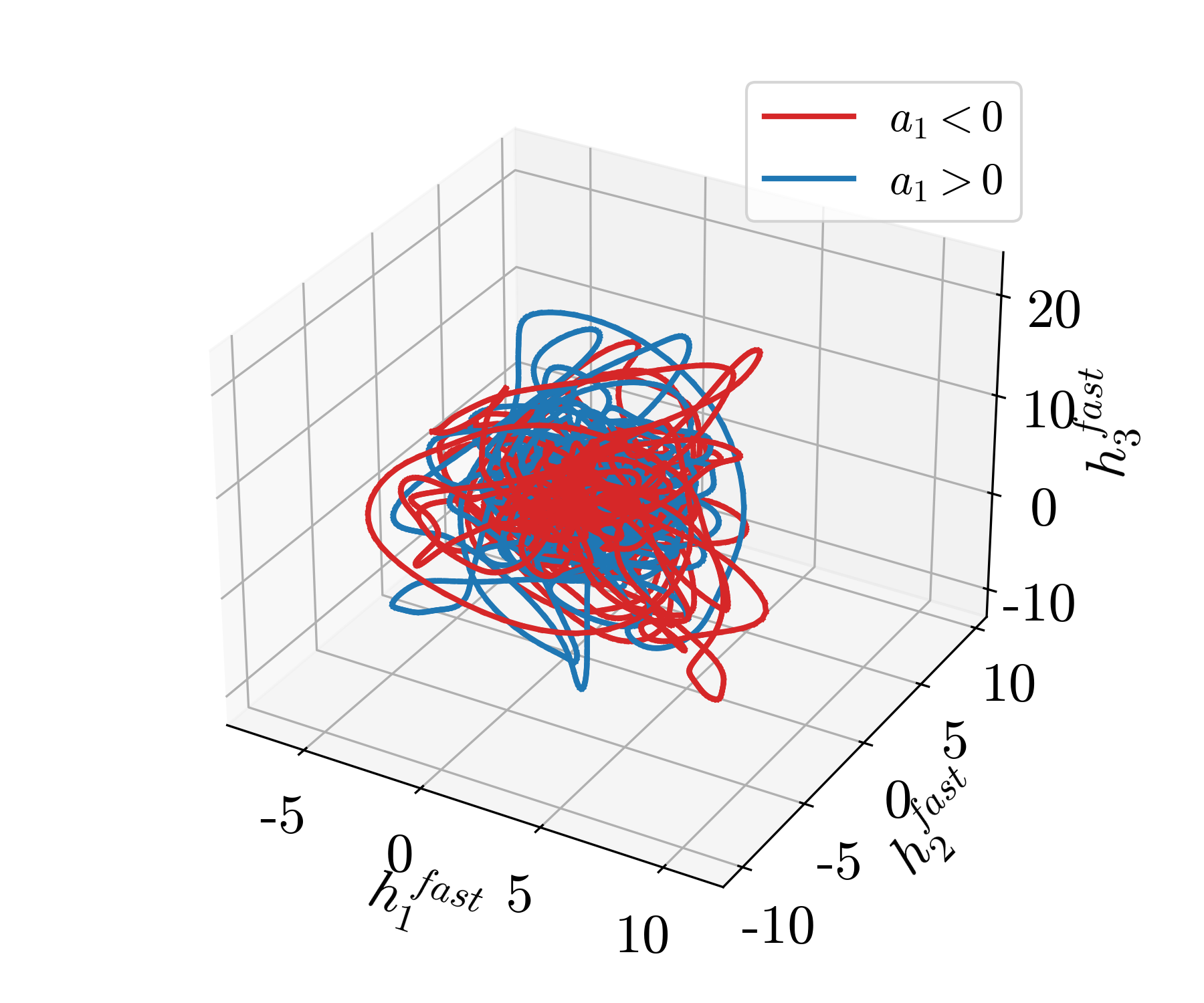} \\
        (b) &
        (c)
    \end{tabular}

    \caption{(a) Comparison of the angular momentum $L_z$ from the DNS, and the different models up to $t=1000$. (b-c) Predicted trajectory of slow, and fast scale latent variables, respectively.}
    \label{fig:am-b50}
\end{figure}

We begin by evaluating the performance  of the proposed  multiscale NODE by   comparing it against two baseline models:  two standard NODE trained on the full latent trajectory using two different   single timescale   resolutions ($\tau=5$ and $\tau=50$). 
Figure~\ref{fig:am-b50}a compares predictions over 1000 time units obtained using these three NODE models which start  at the same initial condition.
Since the reversal cycle dominates the long-term dynamics of RBC, we assess model performance via the angular momentum $L_z(t)$.
The model trained with  $\tau=5$ effectively captures the high frequency fluctuations (fast dynamics), but drifts over time,  leading to poor prediction of flow reversals and large scale behavior. On the other hand, the model trained with $\tau=50$  captures the gross timing of reversal but fails to resolve fast oscillations, missing important features of the turbulent dynamics (e.g., for additional evidence of this limitation,  figure \ref{fig:T-omega-comp} in Appendix  compares the predicted temperature and wall shear rate at two probe locations along the wall). The multiscale model overcomes these limitations;  it accurately reproduces both the long time flow reversals  and the high frequency fluctuations. As a result, it provides a  closer match to the DNS.
These results underscore the importance of explicitly decoupling time scales in latent space modeling and demonstrate the superior performance of the multiscale NODE framework for accurate long time prediction in multiscale chaotic systems. 

Figure \ref{fig:am-b50}b,c show the predicted trajectories of the slow and fast scale latent variables obtained from the multiscale NODE model. Notably, these predicted trajectories exhibit structural features and dynamical behaviors similar to those observed in DNS trajectories shown in figure \ref{fig:traj}. The slow-scale trajectory reproduces the bistable organization observed in the DNS, while the fast-scale dynamics remain consistent with the expected turbulent fluctuations.  This consistency supports that the multiscale NODE model successfully captures the underlying dynamics of the system and preserving its physical interpretability.

\begin{figure}
    \centering
    \includegraphics[width=0.9\textwidth]{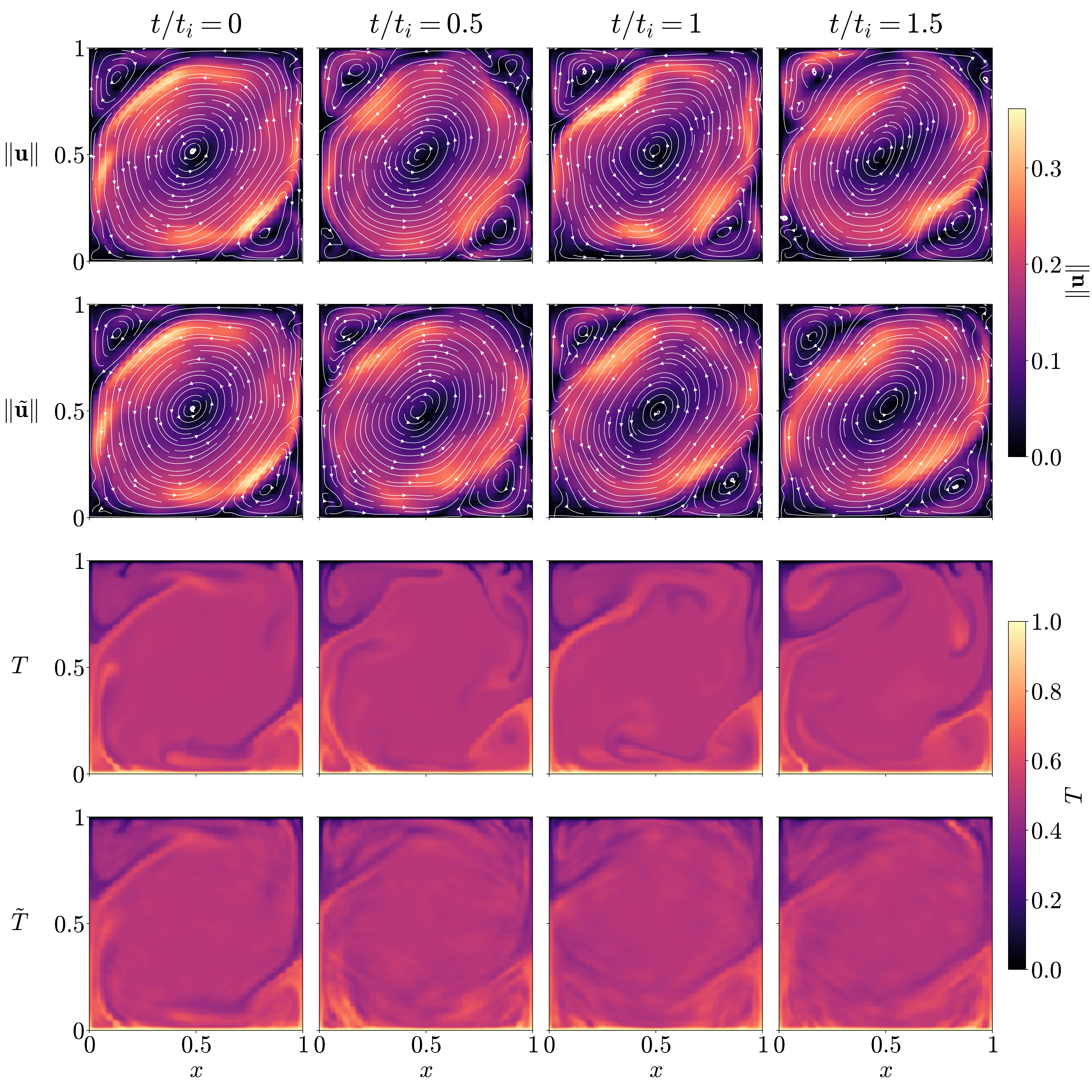}
    \caption{Snapshots of the velocity magnitude $\|\mathbf{u}\|$  and temperature from the DNS and  multiscale NODE for the same IC presented in Figure~6
    Here the correlation time $t_i = 21.2$.}
    \label{fig:field-50}
\end{figure}


Next, having established that the multiscale NODE framework outperforms standard NODEs in capturing multiscale dynamics, we evaluate its performance in reconstructing short-time trajectories.
Figure~\ref{fig:field-50} compares snapshots of the  DNS fields and model-predicted reconstructions evolved from the same initial condition at $t/t_i = [0, 0.5, 1,1.5]$, here $t_i$ is the characteristic integral time $t_i=\int_{0}^{\infty}C(t)dt\approx 21$, where $C$ corresponds to the temporal autocorrelation of the angular momentum $L_z$ (e.g., temporal autocorrelation is defined below). 
In all cases, the multiscale NODE retains key flow features such as vortex cores and thermal plumes. While minor smoothing of filamentary structures is observed, the model consistently preserves the dominant spatial structures, proving its ability of   reconstructing short-term trajectories.

\begin{figure}
    \centering
    \includegraphics[width=0.6\linewidth]{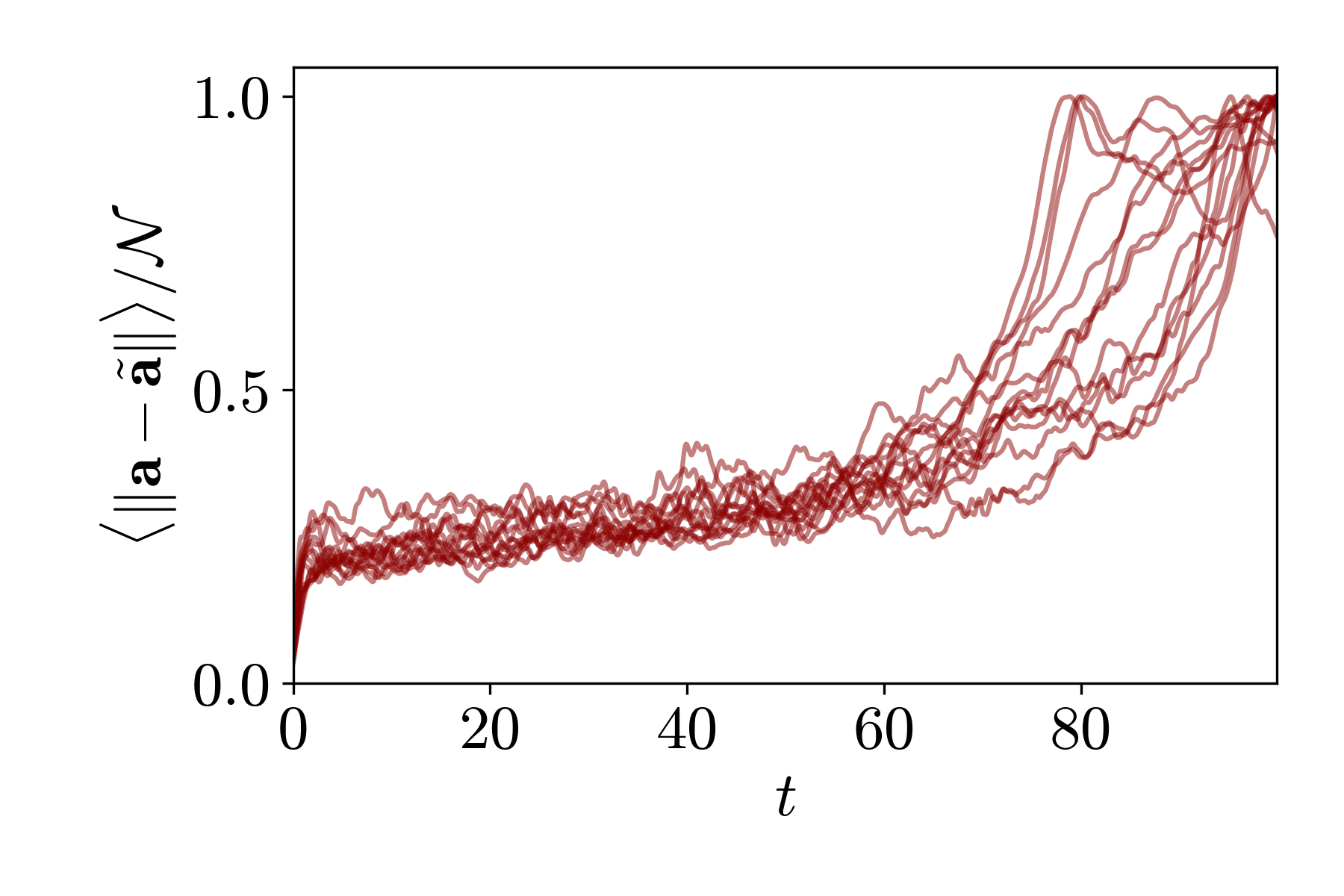}
    \caption{ Normalised short time tracking error for 15 arbitrary initial conditions using multiscale NODE  up to $t = 100$.}
    \label{15ic}
\end{figure}

While the previous result shows predictions from a single initial condition, Figure~\ref{15ic} presents the tracking error for 15 randomly chosen initial conditions. We plot  $\left \| \boldsymbol{a}(t)- \tilde{\boldsymbol{a}}(t) \right \|_2^2/\mathcal{N} $, where $\mathcal{N}$ denotes the maximum error between true solutions and predicted solutions.  
The tracking error initially grows rapidly but soon saturates and fluctuates
 up to $t=60$. This indicates that although the predicted and true trajectories diverge pointwise, they remain on the same reversal cycle within the attractor. The subsequent increase in error beyond $t=60$ likely corresponds to transitions between reversal cycles or other dynamically distinct regions of the attractor, where the model struggles to stay synchronized  with the true trajectory. 

To assess whether the model captures the characteristic time scales of the flow, we analyse the temporal autocorrelations of the instantaneous kinetic and potential energy. The autocorrelation of kinetic energy  reflects the coherence of the flow and the timing of reversals, while that of potential energy  provides insight into the evolution of thermal structures driving these reversals. The  kinetic and potential energies are defined as
\begin{equation}
    E_k(t) = \frac{1}{2} \iint |\boldsymbol{u}(x,y,t)|^2 \, dxdy,
\end{equation}
\begin{equation}
    E_p(t) = -Pr \iint y \, T(x,y,t) \ dxdy.
\end{equation}
Temporal autocorrelation $C_i(\tau)$ (here subscript $i$ corresponds to $k$ or $p$ for  $E_k$ or $E_p$, respectively) is computed as
\begin{equation}
    C_i(\tau) = \frac{\langle (E_i(t) - \bar{E_i}) (E_i(t + \tau) - \bar{E_i}) \rangle}{\langle (E_i(t) - \bar{E_i})^2 \rangle},
\end{equation}
where $\langle \cdot \rangle$ denotes averaging over the entire time window and $\bar{E}$ is the temporal mean of $E_i$.

\begin{figure}
    \centering
\begin{tabular}{cc}
    \includegraphics[width=0.5\linewidth]{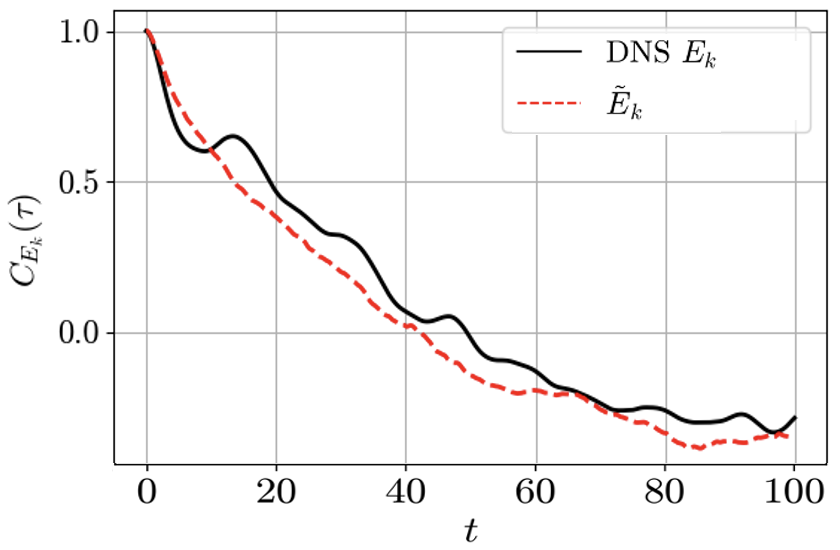}
     & 
    \includegraphics[width=0.51\linewidth]{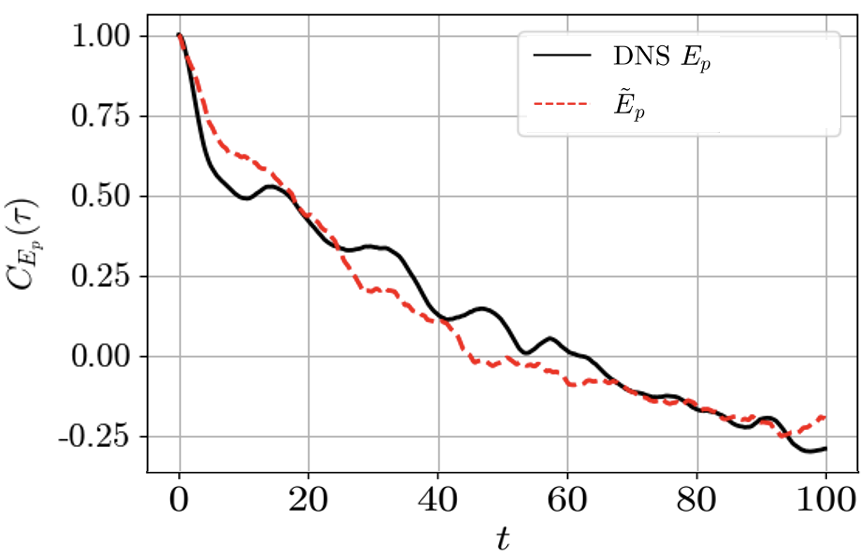}\\
     (a) & (b) \\ 
\end{tabular}
    \caption{ Temporal  autocorrelation of kinetic energy $E_k(t)$ (panel a), and  potential energy $E_p(t)$ (panel b), for the DNS and the model.}
    \label{fig:autocorr}
\end{figure}

Figures~\ref{fig:autocorr}  shows the temporal autocorrelation curves for $E_k$ and $E_p$. 
The autocorrelations of both kinetic and potential energy decay over comparable timescales, with profiles that closely agree between the DNS and the multiscale NODE predictions. This behavior indicates that the multiscale model captures the  temporal dynamics.  Finally, when running long-time series predictions, we observe that the dynamics remain numerically stable. Collectively, these results confirm that the multiscale model retains the essential temporal dynamics of both fast and slow scale components, which is critical for long time forecasting and representation of turbulent convection dynamics.

Next, we will assess the accuracy of the data-driven model near wall boundaries.
In RBC, boundary-layer dynamics play a critical role in controlling both momentum and heat transfer, and inaccuracies in near-wall gradients can substantially impact global system dynamics. To evaluate the fidelity in capturing these localized features, we  examine the wall shear rate and wall temperature. The wall shear rate, $\dot{\gamma}$, is defined as the normal derivative of the tangential velocity at the wall, such as 
\begin{equation}
    \dot{\gamma}(t) = \left. \frac{\partial u_t}{\partial n} \right|_{\text{w}},
\end{equation}
where $u_t$ is the tangential velocity component along the wall, and $n$ denotes the coordinate normal to the  wall.

\begin{figure}
    \centering
    \includegraphics[width=1\textwidth]{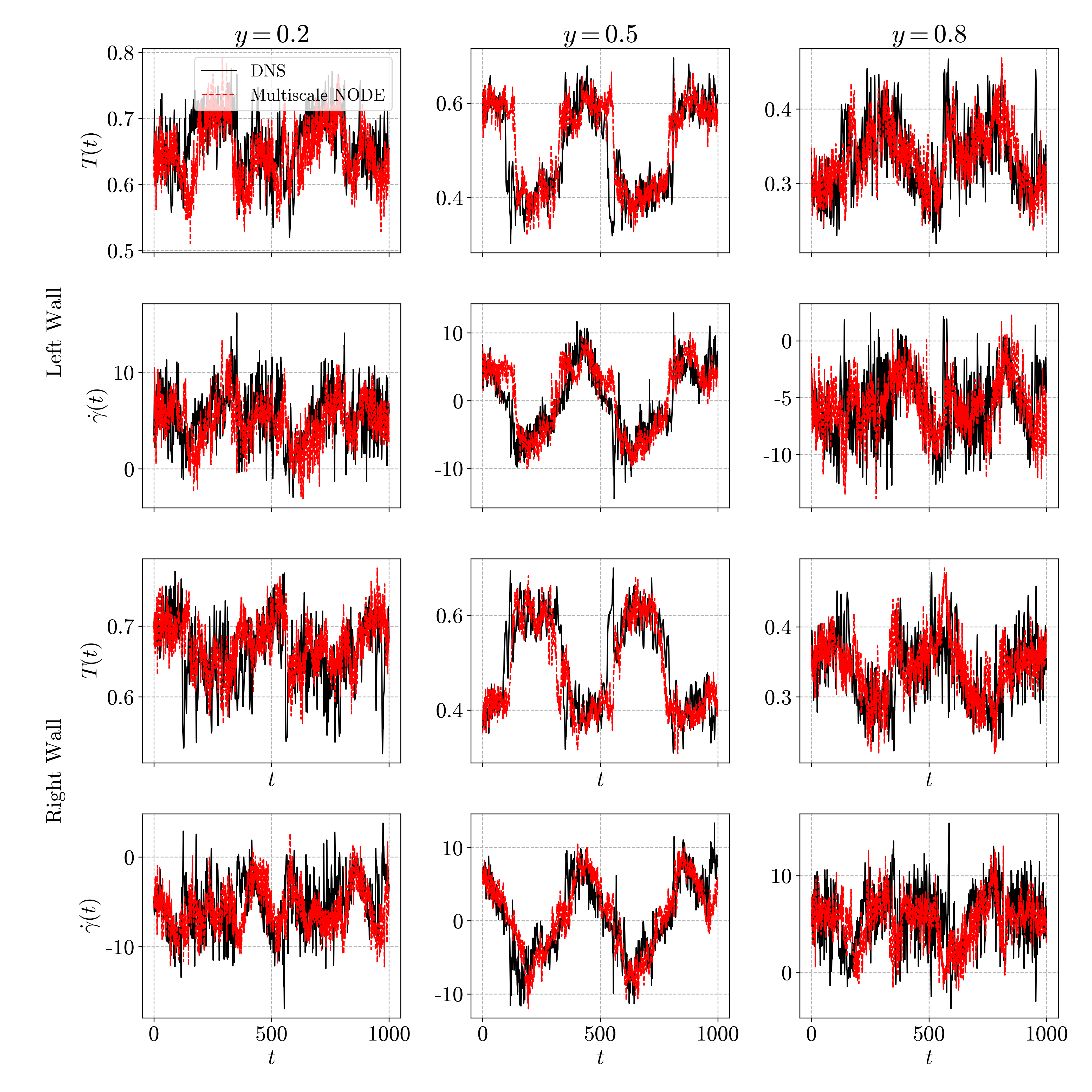}
    \caption{   Time evolution of temperature and shear rate ,$\dot{\gamma}$, on both left and right walls, evaluated at \(y=0.2\), \(0.5\), and \(0.8\).}
    \label{fig:temp-wall}
\end{figure}

Figure ~\ref{fig:temp-wall} shows  time series of the wall temperature and wall shear rate at six  probe locations along the vertical boundaries, positioned at  $y = 0.2$, $0.5$, and $0.8$ on both the left and right walls. In the DNS results, the temperature signals exhibit slow, quasi-periodic oscillations associated with the LSC  reversal events,  during which the hot and cold thermal plumes swap sides of the cell. These reversals induce  shifts in both the mean and variance of the wall temperature. Concurrently, the wall shear rates fluctuate strongly during reversals, reflecting the abrupt changes in near wall velocity gradients as the flow reorients. These reversal dynamics are superimposed on faster, small scale boundary layer fluctuations present throughout the time series. During a reversal,  the central probe clearly captures the transition, as indicated by a change in the sign of the wall shear rate.   The multiscale NODE model exhibits excellent agreement with the DNS signals across all probes, capturing both the low frequency dynamics  and the high frequency boundary-layer fluctuations. This agreement highlights the model’s ability to predict slow global dynamics, such as flow reversals,  while also preserving the rapid variations characteristic of near wall activity.

Finally, to assess the long-term statistical consistency of the NODE model with respect to the DNS, we perform a waiting time analysis focused on reversal events.
Waiting time statistics have been widely used to characterize the temporal structure of rare or intermittent events in  stochastic systems~\citep{Gernert_2014}. These statistics offer insight into whether the reduced-order model  captures not only the frequency but also the temporal correlations of such  transitions. Deviations from exponential waiting time distributions may indicate memory effects or non-Poissonian dynamics—features often linked to underlying physical mechanisms such as intermittency, long-range correlations, or metastability~\citep{Gernert_2014}.
As demonstrated in figure 3b, reversals correspond to the first POD coefficient $a_1(t)$, and to enhance computational efficiency, we therefore identify reversal events as zero crossings of $a_1(t)$ in the following analysis.

For both the DNS  and the NODE predictions, we randomly sample $\mathcal{O}(10^3)$ initial conditions, and predict each one forward in time for $10^3$ time units. The waiting time $\mathcal{T}_i$ is then defined as the interval between two successive reversal events,
    $\mathcal{T}_i = t_{i+1} - t_i$,
where $t_i$ is the time of the $i$-th reversal.
To statistically characterize the sequence $\mathcal{T}_i$, we compute the empirical probability density function (PDF), the survival function, and the cumulative hazard function. The empirical PDF reveals the distribution of waiting times between reversals and highlights the dominant reversal time scales. The survival function, 
$S(t) = \mathbb{P}(\mathcal{T}> t)$,
represents the probability that a waiting time exceeds time $t$. It provides insight into the persistence of the current state before a reversal occurs and is particularly useful for detecting deviations from memoryless behavior.
The survival function
$H(t) = -\log S(t)$,
characterizes the accumulated likelihood of a reversal occurring by time $t$.

\begin{figure}
    \centering
    \begin{tabular}{cc}
        \includegraphics[width=1\textwidth]{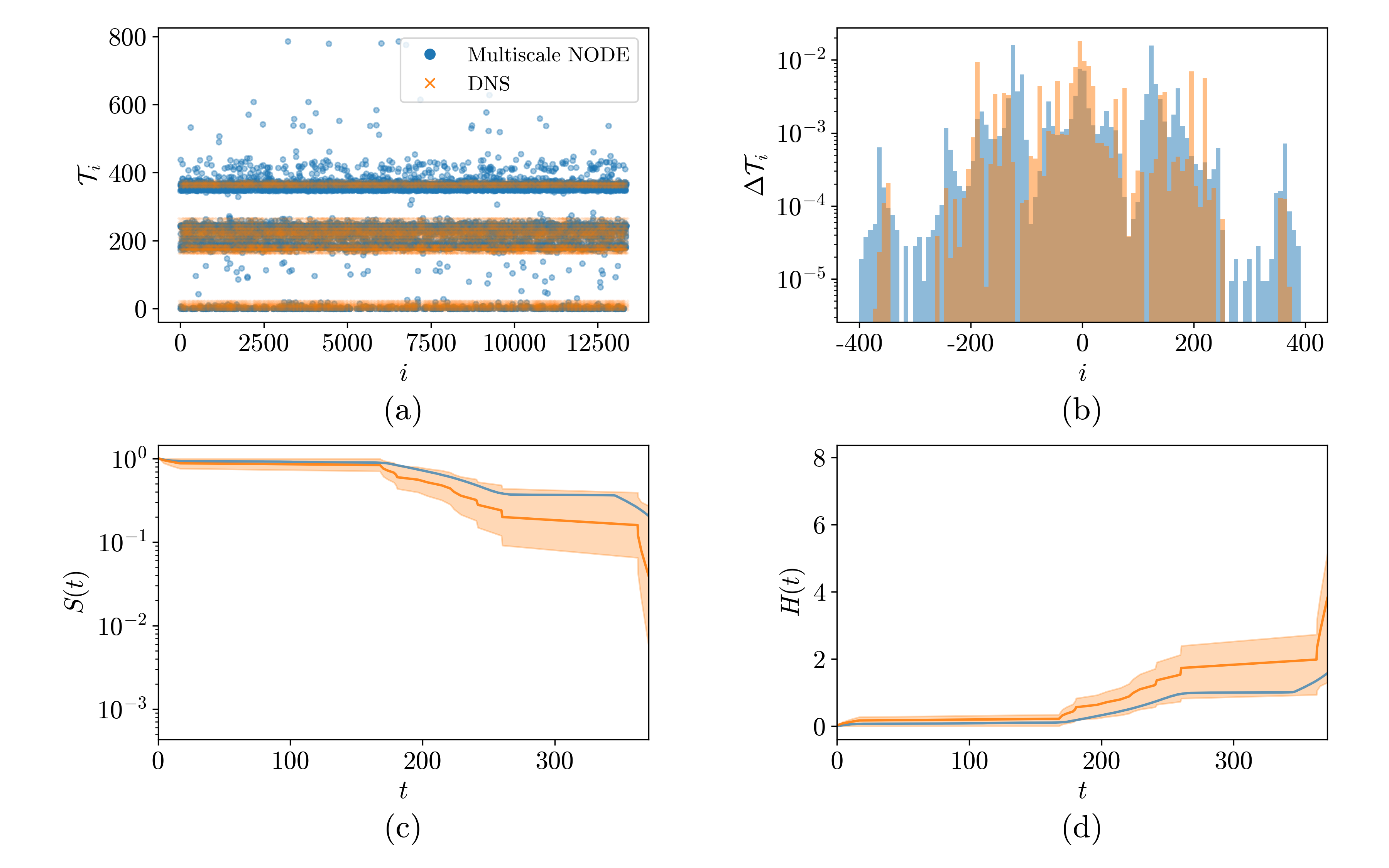}
    \end{tabular}
    \caption{Statistical analysis of waiting time between reversal events from DNS and NODE prediction: (a) waiting time $\mathcal{T}_i$, (b)   distribution of reversal time differences $\Delta \mathcal{T}_i$, (c) survival function and (d) cumulative hazard function.
    }
    \label{fig:waiting_time_analysis}
\end{figure}

Figure~\ref{fig:waiting_time_analysis} compares 
the predicted statistics from the NODE with those from the DNS. Figure~\ref{fig:waiting_time_analysis}a shows the  DNS waiting times are sharply quantized, forming distinct horizontal bands. 
This quantization is likely due to limited DNS sampling, as the training data contains approximately 17 reversal events. In contrast, the NODE model predictions for waiting times are more diffuse and continuous, lacking this temporal quantization. However, the NODE captures certain statistical features remarkably well. In particular, the distribution of reversal time differences \QC{$\Delta \mathcal{T}_i = \mathcal{T}_{i+1}-\mathcal{T}_i$}
(see figure~\ref{fig:waiting_time_analysis}b), agrees closely with that of the DNS, indicating that the NODE 
reproduces the variability of timing between events, even if not the absolute times themselves.
The survival function $S(t)$ further highlights this partial agreement (see figure~\ref{fig:waiting_time_analysis}c). The shaded region represents a $95\%$ confidence interval estimated from the DNS. We observe that the survival function curve for multiscale NODE is  located in the confidence interval, indicating a good agreement in statistical prediction. We also noticed that the model aligns well with the DNS for small values of $t$, but slightly overestimates survival probabilities beyond $t \approx 175$. This is because the model may occasionally predict  long waiting times $\mathcal{T}_i$ (as shown in figure~\ref{fig:waiting_time_analysis}a). Moreover, both the DNS and multiscale NODE survival curves exhibit a similar zigzag pattern, reflecting the effective memory inherent in the reversal dynamics. These observations align with the cumulative hazard function $H(t)$ (see figure~\ref{fig:waiting_time_analysis}d), which   increases nonlinearly, indicating that the likelihood of reversal depends on how long the system has persisted without switching. This consistency supports that the multiscale NODE model can capture both the statistical distribution and the temporal structure of reversal events.

\section{Conclusion \label{conclusion}}

\QC{We have shown that the dynamics of   turbulent RBC with intermittent large-scale circulation reversals can be captured by a  low-dimensional latent evolution when the intrinsic separation of time scales is taken into account. The reduced system reproduces the organization of the instantaneous fields, the structure of second-order statistics, and the long-time evolution of global observables, including the angular momentum associated with the large-scale circulation. In particular, the waiting-time distribution of reversal events is recovered, indicating that the rare-event dynamics are retained in the reduced description.}

\QC{
For the 2D Rayleigh--Bénard convection in a square cell at Rayleigh number $\mathrm{Ra} = 10^8$ and Prandtl number $\mathrm{Pr} = 4.3$, the effective DNS description is reduced from $\mathcal{O}(10^5)$ degrees of freedom to a deterministic latent dynamical system of $\mathcal{O}(10^1)$ variables. The approach combines an initial linear reduction via POD, followed by a nonlinear reduction using multiscale autoencoders. The resulting latent representations are modeled using two separate NODEs. The evolution is integrated autonomously for long times without re-encoding or access to the full state. The comparison with single-stage reductions shows that the principal obstruction to low-dimensional modeling is not only the number of variables but the mixing of processes evolving on widely different time scales. By explicitly separating slow and fast components, the present framework captures both the gradual drift of the large-scale circulation and the rapid fluctuations that trigger the reversal. The reconstructed latent trajectories reproduce the observed temporal asymmetry of the switching process, with a slow build-up followed by a rapid transition, indicating that these events arise from deterministic multiscale interactions.}

\QC{
These results provide empirical evidence that the long-time dynamics of an intermittent turbulent flow can be represented in a compact state-space when its multiscale temporal structure is respected. More generally, they identify time-scale separation as a necessary ingredient for obtaining dynamically self-consistent low-dimensional descriptions of high-dimensional chaotic systems with rare events.
Future work will focus on quantifying the dynamical properties of the latent system and on extending the multiscale construction to three-dimensional convection and wall-bounded turbulence. Additionally, the latent dynamics can also serve as a reduced control environment, in which reinforcement learning can be used to identify control policies in the low-dimensional representation, which will then be transferred to the full space, enabling closed-loop actuation of high-dimensional chaotic systems while preserving the underlying invariant dynamics \cite{chen2025stabilizingrayleighbenardconvectionreinforcement}.
}

\subsection*{Acknowledgments }
This research used the Delta advanced computing and data resource which is supported by the National Science Foundation (award OAC 2005572) and the State of Illinois. The authors gratefully acknowledge useful discussions with Andres Castillo-Castellanos.  \\

 \subsection*{Data availability }

The code and data (in compressed format) used in the paper is available in the group GitHub repository.
The complete dataset is available from the authors upon request.

\section{Appendix}
\subsection{ Neural Network architecture for NODE}

Table \ref{table_arch_node} summarizes the neural network architecture used in the multiscale NODE framework to model the temporal evolution of the coupled slow and fast components. The selected architecture was obtained through trial and error, by varying activation functions, the number of layers, and other hyperparameters.



\begin{table}
\caption{Architecture of stabilized NODE model for latent dynamics}
\label{table_arch_node}
\begin{description}
  \item[Main Network]
  8-layer fully connected network acting on the latent state
  $\boldsymbol{h}$:
  Linear$(d_h,600)$, GELU activations, ending with Linear$(600,d_h)$.

  \item[Linear Map]
  Pre-defined linear operator $\boldsymbol{A}$ acting on
  $\boldsymbol{h}$, implemented as a fixed-weight linear layer
  Linear$(d_h,d_h)$ without bias.

  \item[Output]
  Latent dynamics given by
  $\boldsymbol{g}(\boldsymbol{h})
  = \boldsymbol{A}\boldsymbol{h} + \mathcal{N}(\boldsymbol{h})$.
\end{description}
\end{table}

\subsection{Wall probes predictions for multiscale NODE and standard NODE ($\tau = 50$)}

\begin{figure}
     \centering
     \includegraphics[width=\linewidth]{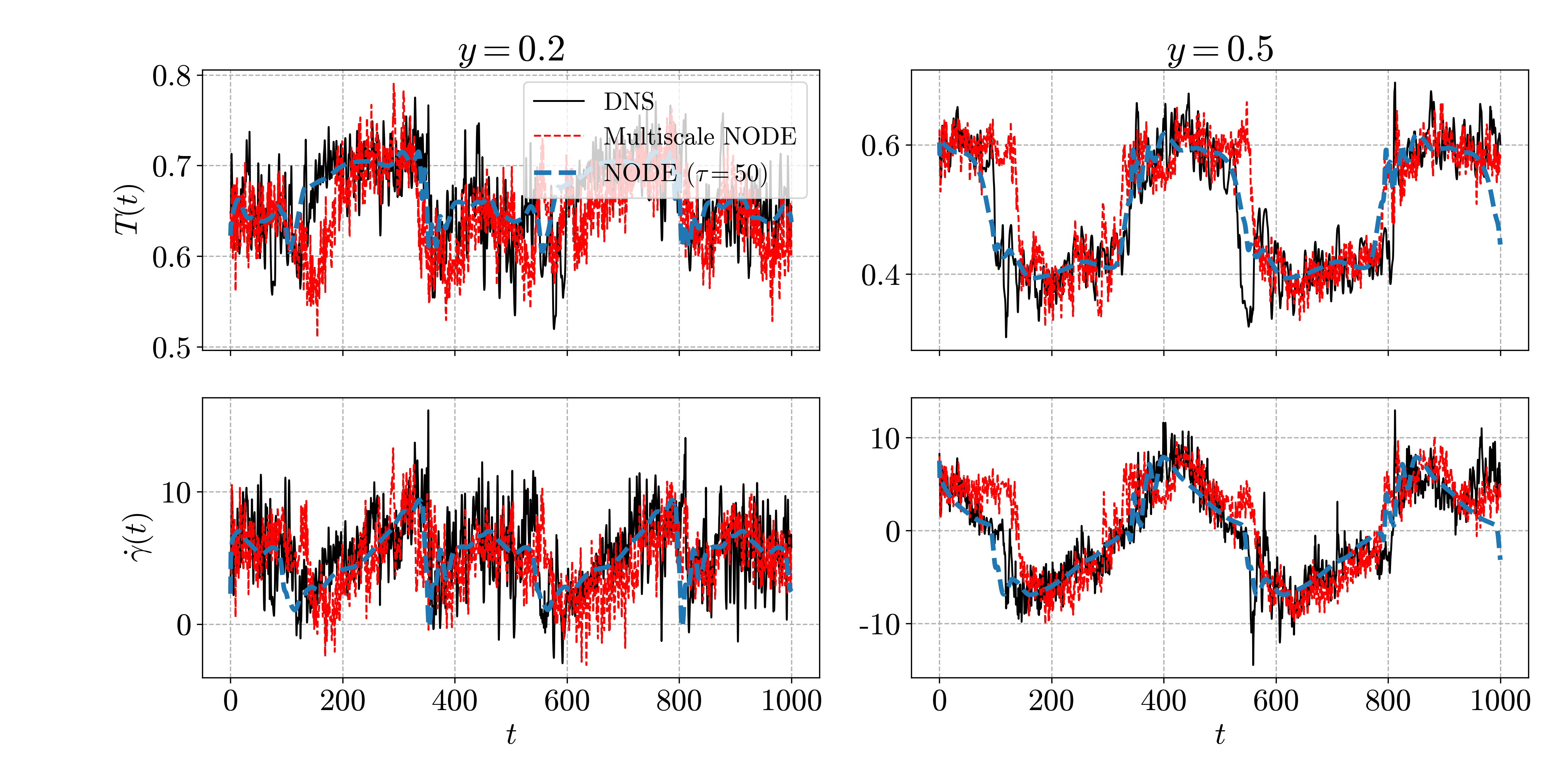}
     \caption{Predicted temperature and shear rate ($\tau$ = 50) at $y = 0.2, 0.5$ along the left wall.}
     \label{fig:T-omega-comp}
\end{figure}

To further highlight the advantage of multiscale NODE over the standard NODE with $\tau = 50$, figure \ref{fig:T-omega-comp} compares the predicted temperature and shear rate at $y=0.2,0.5$ along the left wall. While the standard NODE ($\tau = 50$) is able to capture the general trend of temperature and shear rate, it fails to capture fluctuations, which is associated with chaotic dynamics for this system. In the contrast, the multiscale NODE can capture both the trend and the fluctuation successfully.

\bibliography{apssamp}

\end{document}